\documentclass[letterpaper,12pt, preprint]{article}   
\usepackage{osajnl2} 
\usepackage{epsfig}
\usepackage{subfigure}
\usepackage{graphicx}

\let\~=\tilde

\begin{document}

\title{Imaging X-ray sources at a finite distance in coded mask instruments}

\author{Immacolata Donnarumma,$^{1,*}$  Luigi Pacciani,$^{1}$  Igor
  Lapshov,$^{1}$       
and  Yuri Evangelista$^{1}$}


\address{$^1$  INAF-IASF Roma, Via Fosso del Cavaliere 100, Rome, Italy}  

\address{$^*$Corresponding author: immacolata.donnarumma@iasf-roma.inaf.it}

\begin{abstract} We present a method we developed for the correction of the
  beam divergence in finite distance sources imaging through coded mask instruments. We
discuss the defocusing artifacts induced by the finite distance
showing two different approaches to remove such spurious effects.
We applied our method to one-dimensional coded-mask systems, 
although it is also applicable to 2D systems as well. 
We provide a detailed mathematical description of the adopted
method and of the systematics introduced in the reconstructed
image (e. g. the fraction of source flux collected in the
reconstructed peak counts). The accuracy of this method was tested
by simulating point-like and extended sources at finite distance
with the instrumental set-up of the SuperAGILE
experiment, the one dimensional coded-mask X-ray imager on-board
the AGILE mission. 
We obtained reconstructed images of good quality and high source location accuracy. 
Finally, we show the results obtained by applying this method to real data
collected 
during the calibration campaign of SuperAGILE. Our method was demonstrated to
be a powerful tool to investigate the imaging response of the
experiment, particularly, the absorption due to the materials
intercepting the line of sight of the instrument and the conversion between detector pixel
and sky direction.

\end{abstract}

\section{Introduction}

``Coded mask aperture imaging technique'' is a spatial
multiplexing technique to image X-ray sources over
large fields of view, at X-ray energies where the absorption is
dominated by the photoelectric effect, typically below 100-150
keV. Although the new multi-layer and long focal length telescopes
are now starting being used to focus X-rays between 10 and 70-80 keV, this
can be only done over a very narrow field of view (below 1 degree)
and the coded mask technique is not surpassed yet for imaging large solid angles.

Coded aperture systems encode the sky radiation through a
\textit{mask} of opaque and transparent elements before detector records it. The illuminated mask pattern
projects a ``shadowgram'' on the detection plane shifted with
respect to the central position by an amount that depends on the
direction towards the point source. The source position in the field of
view is reconstructed at a level of accuracy depending on the
angular resolution of the system and the number of counts. The
experimental parameters of a coded mask system can be optimized to
fit the scientific requirements. For example, the
 angular resolution can be optimized at design level by a proper choice of 
the size of mask elements and a mask-detector separation. 
On the other hand coded mask systems have the disadvantage of no
direct imaging that implies the need of decoding procedures to
reconstruct the sky image. Another disadvantage with respect to
direct imaging is that the Poisson noise from any source in the sky is
induced in every other position of the sky. We refer the reader to
\cite{izand94} for an extensive and general discussion
of the design and properties of coded mask systems. Here we limit
our discussion to the properties relevant to the algorithm we
have developed.

The image of a source in the sky is obtained by correlating the
image accumulated by detector with the proper decoding
patterns.
Let $D$ be 
the detector image of the sky, then D is given by the
convolution of the true sky image $S$ with the mask pattern $M$
plus the background $B$:

\begin{equation}D = S \otimes M + B.\end{equation}

It is possible to reconstruct the sky image $S'$ by 
cross correlating the detector image with the reconstruction
matrix  R, which has to be opportunely defined:

\begin{equation}S' = D\odot R= S \otimes (M \odot R) + B \odot R.\end{equation}

The reconstruction is obtained when  $M\odot R$ is the identity
matrix. This can be achieved if the mask is based on a
pseudo-random URA (Uniformly Redundant Array) pattern, built using
cyclic difference sets for which the auto-correlation function
is a $\delta$ function (see \cite{gupo76}). If the mask pattern presents an open fraction of 0.5
(the number of opaque elements equal to the number of the transparent ones) the
reconstruction matrix is given by $R{ij}=1,-1$ for transparent and
opaque elements, respectively. It is evident from Eq. (2)
that this kind of matrix allows for the  background subtraction in the
assumption of its uniformity. Generally speaking, if the complete
mask pattern is projected on the detector then $S'=S$. For several
mask-detector systems this is true only for a limited fraction of
the field of view, while the remaining  sky is only
partially coded due to the collimator shadow. In these regions of the reconstructed images a source induces false peaks (the so-called ``coding noise"). The removal of these
false peaks occurs by using an Iterative Removal of the Sources in the sky (IROS).
We refer the reader to \cite{izand92} for details about
the removal of such effect from sky images.

\indent Imaging source at finite distance (i. e. divergent beams)  
is more complicated because the size of the
shadow in the detector plane is magnified depending on the
distance of the source to the plane itself. In this paper we will call the
effect of the beam divergence on the optical system mask-detector  as a
``defocusing'' effect, although this term is not fully appropriate.   

Using simple geometrical considerations the magnification $m$ of the mask
pattern size is related to the source to the mask plane distance (here $z$) and to the mask-detector separation $f$:

 \begin{equation}m = (z+f)/z. \end{equation}
The standard deconvolution procedure does not allow the source peak
reconstruction because the beam divergence defocuses the incident radiation. 
However we can consider the detector image as the result of
convolution of the source with a virtual mask, i. e. the magnified
version of the true mask, or alternatively we can consider a virtual 
detector (the \textit{reduced } version of the true one) as the convolution with
the true mask. Both approaches
consist in a sort of a ``collimation'' of the modulated radiation thus suppressing the defocusing 
due to the beam divergence. In the
case of a virtual mask, the focusing requires the replacement of
the reconstruction matrix $R$ in Eq. (2) with a matrix $R'$
obtained by expanding R with a factor of $m$ so that $R'{ij}$ will have
values between -1 and 1, defined by interpolation. Instead, the
use of a virtual detector offers the advantage to leave $R$
unchanged. In both cases, the open fraction value ($0.5$) is altered, 
because only a fraction of the illuminated mask pattern is projected on the 
detector. 
 

Thus, there is no strong argument in support of the choice of either procedure. However, in order to have the same configuration of
mask-detector system as for parallel beams (i. e. the same
correspondence between mask element and detector pixel) and to
decode the image with the same reconstruction matrix $R$, we
decided to develop a reconstruction method based on a virtual
detector ``reduced'' by a factor equal to $1/m$.

The reduction procedure (hereafter ``shrinking procedure'') has been checked on the 1-D coded mask system of a
true experiment, SuperAGILE, 
however this procedure, mainly based on simple geometrical considerations, may
also be used for 2-D systems. 

In the next section we present the reduction procedure together  with a
detailed analytical description of the systematics related to the method
for the case of 1-D coded mask-instruments.

In section 3 we show the reliability of the method by means of
computer simulations, assuming the instrumental set-up that was used to
validate our method by means of experimental tests.
The results of these tests are presented in section 4, where     
our method was applied to the ground calibrations of the
SuperAGILE experiment (\cite{fer07}).


\section{The shrinking procedure}
 
The goal of the shrinking procedure is to remove the artifacts in the source
image due to the geometrical magnification of the mask shadowgram.  
As discussed in the previous section, our approach is to 
demagnify the shadowgram in the detector image by reducing the detector
array size.  
Suppose to have a detector of $r$-elements $D$, by \textit{shrinking} we mean  redistributing all the counts,
 by linear interpolation, on a new array $\tilde D$ (``virtual detector'') with dimension $n$
equal to a fraction $1/m$ of the original size (putting the
last $r-n$ elements equal to zero). The perfect suppression of
magnification artifacts requires an accurate knowledge of the magnification factor, 
that is of both source to the mask plane $z$
and detector-mask separation $f$. 
If these quantities are known, the virtual detector image $\tilde D$, corresponding to
the correct value of $m$, is created and then the standard cross-correlation with the
reconstruction matrix $R$ (see Eq. (2)) is applied. 

For a finely segmented detector it is possible to overcome the knowledge of
these quantities by developing  a  procedure that tries to optimize  the
signal to noise ratio of the source image by altering the value of the
magnification factor.  

This is done  generating a set of virtual detector images (each of them
reduced according to  a proper value of  $1/m$, ``the shrinking factor'').

By cross-correlating these detector images with the reconstruction matrix $R$ we generate a set of  ``sky images'', that
show more or less blurred peaks at a given position. 
The shrinking factor corresponding to the highest signal to noise ratio will
represent ``the best approximation'' of the true source image that is also the
best we can do ignoring the detector response. On the other hand  a maximum likelihood approach
can be used to determine e. g. the Point Spread Function of the instrument when our best shrinking factor is found. We do
not enter into the details of this approach because we are focused on the removal of the source
image artifacts, but we refer to \cite{don06} for a description of this procedure.

Right now we consider the signal to noise ratio of the image as a function of
one variable, $m$. This is not true for a realistic detector for which
the ratio between mask and detector elements size varies typically in the range
[1, 3]. Independently on the infinite or finite distance of the source, the
shape of the source peak in the reconstructed images varies depending on the
position in a ``sky'' bin (hereafter ``source phase''). If the source is located in the center of a sky
bin, i. e. each element of the mask
is projected in exactly one detector element, the fraction of detector counts
collected in the peak is $100\%$. Instead, for all the other locations within a
sky bin the total counts will be collected on a wider sky bin range, each bin
scaled to the fraction of mask element projected in detector pixel (\cite{izand92}) .

Thus, we consider the signal to noise ratio of the source image as a
function of one more parameter accounting for the source phase (within the beam
divergence) with respect to which photons are collimated. 
This is expressed by means of the detector pixel intercepted by this
direction towards the source. It is assumed as the center of the shrinking procedure (hereafter ``the shrinking base point" or $i_{shr}$).

In Fig. 1 we show the value of the peak in the sky image as a function of both
the shrinking factor and the shrinking base point. 
The bi-dimensional maximum shape shows an absolute maximum as a
function of the $z$, being a \textit{periodic}
function in the domain of the shrinking base point (see  bottom panel in Fig. 1). While
it is easy to understand the bi-dimensional maximum along z-axis, a periodic
behavior in the shrinking base point domain is less straightforward to interpret. The maximum along the shrinking
factor axis corresponds to the best focusing, and shrinking
with the wrong factor leads  to defocus, thus
loosing an increasing fraction of the peak counts.


The periodic behavior of the maximum as a function of the shrinking
base point position can be explained by simple geometrical considerations
presented below:
a line connecting the source and the edge of a mask element (see
  Fig.2) that also intercepts the edge of a detector pixel, 
determines a local maximum in the shrinking base point domain (see left
panel of Fig. 2, the short-dashed line passing through the mask
element edge represents one of the directions where the
shrinking starts). In fact, shrinking with respect to this point effectively
means collimating the source incident radiation with  respect to the direction
of this line (see right panel in Fig. 2). The possible directions
for which this happens define the periodicity along the shrinking
base point axis. We note that it is more correct to call it a
quasi-periodicity, because the peak local maxima could be slightly different
depending on the mask pattern intercepted by the corresponding directions.  

We now quantify the periodicity in the peak counts, $p$, as
function of $z$ and of the beam divergence. Let $i_{shr}^{'max}$ be a point of local maximum in the shrinking
base point domain connected with the edge $e'$. The place of points $i_{shr}^{max}$ of the local maxima in the shrinking
base point domain is determined by the following relation:

\begin{eqnarray*}
(i_{shr}^{max}-i_{shr}^{'max}) \times  w_{det} = (e - e') \times w_{mask} \times m\\
\mbox{or, introducing $n= (e-e')$ and $p= (i_{shr}^{max}-i_{shr}^{'max})$} \\
  \hspace{0.8cm} p \times w_{det} = n \times w_{mask}  \times (z+f)/z .
\end{eqnarray*}

where $w_{det}$ and $ w_{mask}$ are the detector pixel and the mask element sizes, respectively.
Since $w_{det}$ and $w_{mask}$ are proportional, then $p \propto n \times
(z+f)/z$, thus $p$ is a rational number that
approximates the corresponding magnification factor $m=(z+f)/z$.
When the source lies at a distance equal to a multiple of the mask-detector separation, it is easy to determine
$p$; e. g. for $z=2f$ the periodicity $p$ is equal to 3 pixels in
the detector domain. 
When $z \rightarrow \infty$  (that is, approaches the
infinite distance case) the maximum as a function of the
shrinking base point becomes an almost constant function ($p/n \rightarrow 1$).

We now describe as it is possible to determine either $z$ or $f$ when
the best magnification factor is found. This requires that
one of the two parameters is known with a reasonable accuracy given the
degeneracy present in Eq. (3).  
For example,  if $z$ is ignored, it will be known within an uncertainty equal to $\Delta f/(m-1)$ ($m=1$ is
reached for $z \rightarrow \infty$)  where
$\Delta f$ includes both the uncertainty introduced by the accuracy of the 
metrological measurement and the error due to its dependence on energy
($\Delta f(E)$). The efficiency of the detector and the operating energy band
require different approaches in the use of shrinking procedure.

Depending on the absorption length of the incident photons, it is possible to
have a large spread in the mask-detector separation, causing a systematic error on $f$.
If the radiation is preferentially absorbed at a given depth in the detector
thickness, one single value of $f$ has to be used with an uncertainty
associated with the dispersion in the absorption. In all the other cases we suggest to use the
shrinking procedure separately for each value of $f$, taking into
account its own error ($\Delta f(E)$).

When $z$ has been determined, it is easy to determine the lateral
displacement of the source $y$ by means of the shrinking base point $i_{shr}^{max}$ corresponding
to a given direction  $\phi$.  It is defined as $y= (z+f) \times tan(\phi)+ i_{shr}^{max}\times w_{det}$ ($w_{det}$
is the detector element size).    

The goal of our method is the deblurring of the source image, that is
approximate the infinite distance case. However, even in the best case, the
result of our method will show a clear difference with the infinite distance
case, due to the systematics introduced by the linear interpolation.

We now quantify the systematics. To this purpose we  compare the sky image of
a source at infinite distance with that
obtained for a point-like source at finite distance, neglecting the source spatial extension.   

Assume a source at on-axis position, that is we neglect the 
flux attenuation correlated with the source phase. 
As opposite to the infinite distance case, where the sky image peak has to
include almost the $100\%$  of the detector recorded counts, in the finite
distance case only a fraction of the recorded counts will contribute to the
peak in the sky image, by an amount that
depends on the distance $z$ and on the beam divergence.\\
\\
Let be $D$ a function describing the counts recorded by the detector pixels.
Shrinking the detector means to build a new function $\tilde D$ that satisfies
the relation $\~D(i)=D(i \times  m)$. $D(i\times m)$ is obtained linearly interpolating between 
the counts of the pixels $[i\times m]$ and  $[i\times m]+1$
($D([i  \times  m]+1)$ and $D[i \times m]$, respectively) where the square brackets indicate the integer part of the number. This  implies that
\begin{equation}
\~D(i)= (i \times m-[i \times m]) \times (D([i \times m]+1)- D[i \times m]) +D[i \times m].  
\end{equation}

For simplicity we define $j=[i\times m]$.
As it is shown in Fig. 3 the counts  $D(j)$ recorded by the j-th detector pixel
depend on the value of the 3 elements of the mask pattern $M( j-1), M(j), M(j+1)$
and on the shadowgram produced in the detector. This means that it also depends on
the fraction of the flux lost by $D(j-1), D(j), D(j+1)$ ($a,b, c$ in Fig. 3),
expressed as a function of the angle $\theta$ in the coding direction (see Fig. 3). 
Then the same dependence applies to the counts of the virtual detector $\~D$. 

Each element of the mask pattern can have two possible values
(0,1), causing $2^{3}$ different combinations of the
numerical terns to exist. The virtual detector pixels will then collect a
number of counts depending on the distribution of the illuminated numerical
terns and on $\theta$ (see Appendix for details).

Finally, a standard cross correlation between mask pattern and
virtual detector can be performed, multiplying by -1 the detector counts in the i-th pixel coming from a tern with
central element equal to 0, $\~D_{i}^{0}$,  and by 1 the counts of detector
corresponding to a tern with central element equal to 1,  $\~D_{i}^{1}$.

In the assumption  that the eight numerical terns are equally
distributed (if not the following relation is still valid), we can calculate the total number of counts on the virtual detector
 by integrating Eq. (4) from the minimum up to the maximum angular divergence
for a fixed $z$; the integration in $\theta$ is done over the values of
$\theta$ corresponding to all the illuminated portions of the detector.

Then the resulting normalized sky image will be:

\begin{equation}
Sky_{norm}= 1 - 2 \times \~D_{int}^{0}/(\~D_{int}^{1}+ \~D_{int}^{0})
\end{equation}
where $\tilde D_{int}^{1} = \displaystyle\int_{\triangle\theta} \tilde D^{1}\times f(\theta) d\theta$, $\tilde
D_{int}^{0}=\displaystyle\int_{\triangle\theta} \tilde D^{0} \times f(\theta) d\theta$ and
$f(\theta)$ is a function accounting for the $\theta$ variation within the
beam divergence.

Eq. (5) tells us how much the ratio between peak counts and the total detector counts is affected by the magnification.
In particular, it easy to see from Eq. (5) and the tern values described in the Appendix that this
fraction increases with $z$ approaching $1$ when the source is at infinite distance.   

There may be cases for which the dependence on $z$ is negligible.
In a mask-detector system, with the ratio between mask element and detector element size equal to 2,  the peak to total counts ratio
is expected to not vary with $z$. In this case, the number of terns to be
considered is reduced from $2^3$ to $4$ (only terns with $2$ equal contiguous
elements  have to be considered) and it is possible to verify that Eq. (5) is constant with $z$.  

With respect to the overall shape of the reconstructed sky image, 
we expect the point spread function to be symmetric because the brightest peak
is obtained when the element of the mask pattern element is in phase with one
detector strip. A small asymmetry may be present due to the
computational accuracy and can be neglected. Consequently we consider any
asymmetry in the reconstructed point spread function as associated with real physical and
geometrical effects in the instrument.\\

In the next sections we describe the results achieved applying the
shrinking procedure to the imaging with the SuperAGILE experiment with
simulations and experimental tests.

\section{Computer simulations of beam divergence effects: SuperAGILE}

To verify the reliability of our method we performed Monte Carlo
simulations of a large set of sources imaged by SuperAGILE. 
\\
SuperAGILE is the hard X-ray monitor on-board AGILE
(Astro rivelatore Gamma a Immagini LEggero, see \cite{tav06}) operating in the nominal energy range 18-60 keV.
It consists of four one-dimensional coded-mask detectors, encoding the same
direction in pairs. The collimator and the mask define a single
(one-dimensional) field of view of $107^{o} \times 68^{o}$ for each of the $4$
detectors, orientated at $90^{o}$ for the two pairs of units.

The masks code is a Hadamard sequence of 787
elements generated with the method of quadratic residue, to which
one element has been added to reach the complete symmetry and
reversibility of the mask. Each nominal mask element is 242 $\mu$m
wide, twice the dimension of the strips of the silicon micro-strip
detector, to ensure the proper sky image reconstruction. For a
detailed description of the experiment we refer the reader to
\cite{fer07} and to references therein.

We simulated  both point-like sources as well as extended sources,
varying their positions in $z$ (source mask-plane distance) and
along the plane parallel to the mask to study the accuracy of the
focusing method as a function of distance and lateral
displacement. The choice of distances from the detector plane is
influenced by the aims to be achieved: as an example, tomography is best achieved
by imaging source located as close as possible; 
understanding detector response to infinite distance
sources requires going far from the detection plane
($z\rightarrow \infty$ is equivalent to $m\rightarrow 1$). Imaging
sources very close to the detection plane gives rise to important
side effects. The radiation intensity reaching the detector is no
longer uniformly distributed because its path-length $r'$ increases
going far from the normal vector $r$ (``inverse square effect").
Moreover, the angular subtense of each hole decreases across the
aperture (``obliquity effect") leading to flux attenuation
proportional to the same angle. We refer the reader to \cite{cafe79}
for details about the inverse square and obliquity effects.

Since we are interested to reach the best focusing of the image,
we choose distances for which the side effects
can be neglected.  Thus  we varied the distance  of the simulated source to the detector plane
in the range $100-300$ cm
(for SuperAGILE, with its mask-detector distance of $\sim 14.2$ cm,
this means from $m\sim 1.1$ to $m\sim 1.05$). On the basis of the reconstructed images and properties such as the peak counts, the
point spread function widening and systematic asymmetry of the shape,
we found no strong arguments to discriminate among these distances (see columns 1,2 in Table 1). 

In particular, the reduction
of the peak counts is $\sim$ 17$\%$ with respect to the source counts 
recorded on the detector  in the range $100-300$ cm. This is what we expected from the
analytical description of the peak-counts attenuation, because for SuperAGILE the ratio between mask elements and detector element size is
equal to $2$  (see previous section). As expected, for a source located at infinite distance 
we did not find any reduction of the peak counts.

The imaging at finite distance also affects the Point Spread Function (see dot line in Fig. 4 for comparison with the infinite distance simulation). A widening of two side lobes (with $\sim10 \%$ of the recorded counts) appears. 
The flux attenuation (or the image widening) becomes larger if one
considers an extended source: for a monochromatic source at 32
keV, with a disk shape 5 mm in diameter we predict a factor of
$\sim$ 40$\%$ (see dashed line in Fig. 4). It is clear that this further widening depends on
the spatial profile of the source.

In the left panel of Fig. 5 we  show the image of a simulated
point source at $z+f=250$ cm obtained
with the standard deconvolution method (see Eq. (2)),
i. e. without applying the focusing procedure. The right panel of
the same figure shows the significant peak emerging after the
shrinking procedure is applied.

By means of these simulations we also studied the error in the source direction
parallel to the optical axis associated with
statistical fluctuations. We found that the
1-$\sigma$ error for the parallel direction $z+f$ is $\simeq 0.1
$ cm if a signal to noise ratio of $\sim  300$ is achieved.

The performed simulations also drove set-up of
calibrations of the SuperAGILE experiment with radioactive sources
in laboratory, given the reconstruction accuracy described
above. In particular the choice of the vertical displacement of
the source was based on a trade-off between the accuracy on
determination of the physical quantities to be calibrated and the
integration time needed for these measurements.

For example one of our major goals was to determine the
mask-detector separation $f$ from the
magnification factor, knowing $z$ (see Eq. (3)), thus
calibrating the relation between  the detector pixel \textit{i} and sky
direction $\theta_{sky}^i$ by means of the following relation:
\begin{eqnarray}
\theta_{sky}^{i} = arctg(i \times w_{det}/f).
\end{eqnarray}



In this case the farther the point source is the better
the accuracy and the determination of $f$ can be. In particular, assuming a $z+f$
distance of $\sim 200$ cm and taking a reasonable accuracy in the
experimental measurement of $z$ into account ($\sim 0.1$ cm), we predict an
accuracy of $\sim$75 $\mu$m in the estimation of $f$ (see Table 1, it is
$\sim165$ $\mu$m at $z+f=100$ cm). 

We now see as this error is propagating on the reconstruction of the sky direction (see Eq. (6)). 
The uncertainty in the
reconstruction of the sky direction as a function of $f$ is given
by  $1/2\times sin(2\theta_{sky}^{i})\times \Delta f/f$, then a $\Delta f$ equal
to 75 $\mu$m corresponds for a 30 degree off-axis position to an
accuracy of 0.013 degrees, while $\Delta f= 165$ $\mu$m
corresponds to 0.028 degrees. These numbers have to be compared
with the value of $\sim$ 0.036 degrees for  sky pixel size  at 30
degrees.
Thus, a distance greater than 200 cm allows to calibrate
the sky direction reconstruction capability with an uncertainty
that is well below the pixel size.

We also estimated the error in the direction orthogonal to the optical axis,
i. e. the lateral displacement of
the source, defined as
follows

\begin{equation}
y = (z+f) \times tg(\phi) + i^{max}_{shr} \times w_{det} +\bigtriangleup
\end{equation}

(where $\phi$ indicates the angle chosen for the shrinking, $i^{max}_{shr}$ is the point of local maximum along the shrinking base point axis, $w_{det}$  is the
detector strip length and finally $\bigtriangleup$ is a correction parameter related
to the detector-mask alignment) with Montecarlo we found an accuracy of
$\vert\Delta$y$\vert  \simeq 0.0045$ cm for a source located at $200$ cm, due to
statistics. This means that for distances larger than 200
cm from the detection plane, we would expect an uncertainty of $y$ less than $3\%$ of detector pixel size.

Therefore, for the ground calibrations of the SuperAGILE
experiment we requested a source distance equal or greater than
$200$ cm. However all the previous discussions assume that we
preserve the counts statistics at any distance, and this is of
course not true, for a given intensity of the calibrating source.
This suggests to not move too far from the 200 cm, in order to
avoid divergence of the integration time needed to reach a given
statistical accuracy.

\section{Laboratory results}

The shrinking procedure was demonstrated to be a powerful tool to
use diverging beam sources at finite distance  to calibrate coded-mask systems. In this section we present how the shrinking
procedure has been used during the calibrations of the SuperAGILE
experiment, currently flying on-board the AGILE space mission (see
\cite{don06} and \cite{eva06} for
the first calibrations of this instrument performed in 2005).

The final SuperAGILE ground calibrations were performed on January
2007. The  experimental setup was studied to locate the
radioactive sources  at $\sim$ 200 cm from the detection plane.
The micrometric position of the sources
was targeted with a laser tracker with respect to reference points
on the spacecraft structure, and to the reference optical cubes
aligned to the stars sensors of the AGILE satellite. For this
calibrations campaign we used $3$ radioactive sources: $Cd^{109}$
(with a main line at 22 keV), $I^{125}$ (line complex at 27-32
keV) and $Am^{241}$ (main line at 59.5 keV). As discussed above,
the main goal of the shrinking procedure is to allow the study of
the instrument imaging response without using parallel beam
sources, and using instead the more easily accessible (quasi)
point-like commercial radioactive sources.
This implied a large set of measurements covering from one edge of
the field of view (FOV) to the opposite one in coding direction as
well as in no coding one. The sample of the scanned positions
covered the range $-50$ to $+50$ degrees with a step of 10 degrees.
The complete scanning was performed only for the $Cd^{109}$. The
other sources were instead used to sample fewer positions in the
center of the FOV (see \cite{fer07}).

In Fig.  6 we present a sample of the zoomed source images for SuperAGILE
detector units. In Fig. 7 we show the full source sky image in the Field of
View for $0$ and $30$ degrees off-axis.  
The quality (quite-symmetric) and the widening 
of the peak shape agree with what is expected from Montecarlo simulations of
a source with similar spectral energy and spatial profile. This is corroborated by    
the values of the distance $z+f$ inferred by the magnification
factors obtained for each measurement. They are in agreement (within the
uncertainties of $\sim1-2$ mm) with the distances measured with
the laser tracker.

Then we can say that the correction works well for each
of the off-axis positions showing as it is influenced only by the
statistics (similar in all measurements). 
Looking at the last two images, at $\sim 40, 50$ degrees, the
correction appears to work fine, as demonstrated by the symmetric
shape of the point spread function, but a distortion of the
baseline is also found. A deeper analysis of this feature showed
the presence of a small ``leak" in the collimator shielding
affecting the image of this detection unit for angles greater than
$20$ degrees.
\\

Given the level of accuracy achieved in the corrected source images, 
it has been possible to use the detector
images reduced by the correct magnification factors as well as
the reconstructed sky images for calibration purposes. In particular
we used the first ones to estimate the transparency of the
materials intercepting the line of sight of SuperAGILE. On the
other hand the sky images has been used to calibrate the pixel-angle
conversion with the purpose to obtain an analytical law.  

This has been done associating the centroid position of the reconstructed
images with the angles provided by metrological measurements (triangle
points in Fig. 8). The simplest relation that connects pixel and angle is
provided by Eq. (6). However we found that a best fit to the data is found
by adding a geometrical offset between mask and detector planes plus a pixel offset $\Delta$ to Eq. (6).   
 
In fact, a better fit
to the data (solid line in Fig. 8) has been obtained using the
following parametric function:

\begin{equation}
\theta^{i}_{sky}=arctg[(m_{st} + (i+\bigtriangleup) \times w_{det})/f]
\end{equation}

where $m_{st}$, $w_{det}$ and $f$  are the distance of
the first mask element in the SuperAGILE reference system, the
size of pixel and the mask-detector separation, respectively. 
As evident in residuals (bottom panels in Fig.8) the agreement
between measured angles and those inferred by the best fit is
quite good, being $\leq 2$ arc minutes (remind for comparison that SuperAGILE
spatial resolution is 6 arc minutes).


We stress as the understanding of such effect is crucial in the
procedure of Iterative Removal of Sources (IROS) in coded mask
instrument. In fact this requires a good understanding of the
instrumental response to a source in a given position in the sky.

\section{Conclusions}

We described the method developed to correct for the beam
divergence in coded mask imaging of sources placed at finite distance from the
detector plane.
The method has been studied and applied on 1-D coded-mask instrument, 
although its extension to the 2-D instruments is feasible.
It works to collimate the modulated radiation from the
X-ray source with reference to one direction within the beam
divergence. The choice of the direction is determined by the best
suppression of defocusing effects of the image, associated to the
magnification of the source shadowgram in the detector.

This requires the knowledge of the source distance to the mask plane and the
mask-detector separation. If these two values
are not known or known within some uncertainty, the best focusing
may be found searching for the highest signal to noise ratio in the
reconstructed image. This is a function of both the source distance to the
detector plane (and then of the magnification factor) and the direction (within the beam divergence) with respect to
which photons will be collimated. We argued that there is only a value of the
magnification factor corresponding to the maximum of the signal to noise
ratio, while  there are several (depending on the distance) favorite
directions to collimate photons, going to one possibility for the infinite
distance case. A detailed study of the
systematics introduced by our method on the reconstructed source image has
been done. We found a reduction of the peak counts compared with the
infinite distance, which amount depends on the source distance and the
beam divergence (neglecting the source spatial extension).

To check the reliability of the method we performed a set of Monte
Carlo simulations of X-rays sources imaged by the SuperAGILE experiment. 

We simulated point-like sources as well as extended sources at finite distance.
We found that the flux attenuation with respect to the parallel beam case mainly depends on the geometric extension
of the source. The dependence on the source distance is quite negligible due
to the particular configuration of the coded-mask system of SuperAGILE.

Then we found  a flux attenuation of $\sim 17\%$ compared with the
parallel beam case for a point-like source. This clearly quantifies the
widening of the Point Spread Function due to the systematics of our method. The flux reduction increases  to $\sim 40\%$ if e. g. a source with a disk shape of $5$ mm in diameter is considered.

By means of these simulations, we estimated the statistical errors on both
orthogonal and lateral displacements of the source obtained with our
method. For a signal to noise ratio of  $~ 300$, we find that the 1-sigma error for the orthogonal displacement is  0.1 cm,
while it is negligible  for the lateral displacement. These
estimations of the method accuracy have been used to improve the ground
calibration set-up of SuperAGILE. Searching for a
good trade-off between the integration time (at the required $S/N$
ratio) and the physical quantities needed to be calibrated, we
fixed a distance of $200$ cm of the source to the detector plane.
We imaged Cd$^{109}$ and I$^{125}$ at $\sim 200$ cm. In
particular, tests with Cd$^{109}$ were done from -50 up to 50 degrees with a step of 10
degrees in the
experiment field of view. In this case we succeeded to obtain a
good reconstruction of the source peak from -40 up to 50 degrees off-axis.
The values of source distance from the detection plane as well as
its angular displacement inferred from the shrinking procedure are
in good agreement with the metrological measurements performed
with a high precision laser tracker ($\sim 100 \mu$m accuracy).
This allowed us to calibrate the conversion between detector pixel
and sky degree within 2 arc minutes uncertainty (to be compared with
SuperAGILE pixel size of 6 arc minutes) and to study the
transparency of materials surrounding SuperAGILE.



\section*{Acknowledgments}

We are very grateful to the SuperAGILE colleagues, to the AGILE
Team and to the Italian Space Agency  providing a continuous
support to the SuperAGILE project. 
In particular we thank Marco
Feroci for his advices about this work and a careful reading of
the manuscript.

\section*{APPENDIX}

We describe as the virtual detector, $\tilde D$, is obtained from the counts recorded by
each of the pixel of the real detector $D$.
As described in section 2. the $i$-th pixel of $\tilde D$ will collect a
number of counts obtained by interpolating between the counts recorded by
$[i\times m]$ and $[i\times m]+1$ pixels. We indicate them with $D(j)$ and
$D(j+1)$ where $j=[i\times m]$.
As it is shown in Fig. 3 they strongly depend on the projected mask tern
$M(j-1), M(j), M(j+1)$: 
 
\begin{eqnarray}
D(j-1)& = & F(M_{j-1})- \alpha  \times  F(M_{j-1})
\\
D(j) & = & F(M_j)- \beta  \times  F(M_{j})+ \alpha  \times F(M_{j-1}) 
\\
D(j+1) & = & F(M_{j+1})-\gamma \times F(M_{j+1})+ \beta \times F(M_j)   
\end{eqnarray}

where the function $F$ is the ``transfer function'' of the mask, which can
assume $0$ or $1$ values. $a, b, c$ are the fraction of the
source flux ``lost'' by $ D(j-1), D(j), D(j+1)$, respectively and are so defined:

\begin{eqnarray*}
 a & = &          (m-1) + \frac{f}{w_{det}} \times tg(\theta) \\
 b & = & 2 \times (m-1) + \frac{f}{w_{det}} \times tg(\theta) \\
 c & = & 3 \times (m-1) + \frac{f}{w_{det}} \times tg(\theta)  \\
\end{eqnarray*}
where $\theta$ indicates the angle in the coding direction.
The Eq. (10), (11) have to be used when $b\le 1$.  

When $b$ becomes greater than 1, that is 
$arctan[(1-2\times (m-1))\times w_{det}/f]< \theta < arctan[w_{det}/f]$,   
$D(j)$ and $D(j+1)$ are expressed as follows: 

\begin{eqnarray}
D(j) & = & \alpha  \times F(M_{j-1})\\
D(j+1)&  = & \beta \times F(M_j) .
\end{eqnarray}

We now show the whole mathematical description  for $b < 1$ although  we will
later discuss the impact of $b \geq 1$ on the general method. 

Substituting  $D(j+1)$ and $D(j)$ described in Eq. (10) and Eq. (11) inside the
Eq. (4), we obtain:

\begin{center}

$ \~ D(i)= (m-1) \times [F(M_{j+1})- \gamma  \times F(M_{j+1})+ \beta \times
F(M_j)-(F(M_j)+ \alpha \times F(M_{j-1})-\beta \times F(M_{j}))]+F(M_j)-\beta
\times F(M_{j})+ \alpha \times F(M_{j-1})$
\end{center}

Since each element of the mask pattern could assume two possible values (0,1), we have to distinguish among $2^{3}$ different combinations of the numerical tern.
Then  the $i$-th pixel of the virtual detector $\tilde D$ will collect a number of
counts among  the following possibilities:
\newpage
\begin{eqnarray*}
\mbox{Tern values} &    &   \tilde D(i) \\
111        &      &  \frac{1}{m} \\
000        &      &   0  \\
100        &      &    [\frac{(m-1)}{m}+ \frac{f}{w_{det}} \times \frac{tg(\theta)}{m}]\times (2-m)  \\
010        &      &     (m-1)\times[4\times\frac{(m-1)}{m+2} \times \frac{f}{w_{det}} \times \frac{tg(\theta)}{m}]+ \frac{(2-m)}{m} + \frac{f}{w_{det}} \times \frac{tg(\theta)}{m}\\
001        &      & (m-1) \times [\frac{(3-2 \times m)}{m} +\frac{f}{w_{det}}     \times  \frac{tg(\theta)}{m}] \\
110        &      &  (m-1) \times [\frac{(2 \times m-3)}{m}+   \frac{f}{w_{det}}     \times  \frac{tg(\theta)}{m}]+\frac{1}{m}\\
011        &      &  (m-1) \times [\frac{(m-1)}{m}+\frac{f}{w_{det}} \times \frac{tg(\theta)}{m}]+\frac{(2-m)}{m}-\frac{f}{w_{det}}\times \frac{tg(\theta)}{m}\\
101        &      &  (m-1)\times[\frac{(-3 \times m+4)}{m-2} \times
\frac{f}{w_{det}} \times \frac{tg(\theta)}{m}]+\frac{(m-1)}{m}+\frac{f}{w_{det}} \times \frac{tg(\theta)}{m} 
\end{eqnarray*}

These expressions define the integrand functions in  Eq. (5).
If $\Delta \theta$ is the integration domain we remind that when 
$arctan[(1-2\times (m-1))\times w_{det}/f]< \theta < arctan[w_{det}/f]$ the
integrand function  has to change according to Eq. (12), (13). 
However we stress that this range of $\theta$ is $2\times(m-1)$ wide 
and the contribution of the new integrand function is proportional to $m-1$, that
implies a contribution $\propto (m-1)^2$ to the overall integral.
It is then clear that when the source is placed at a distance for which $m<
1.1$, this additional term could be neglected (being of order of few percent).

We note that this method is valid when the integration over
$\theta$ is possible.  This means that the $\triangle\theta$ integration range
must be large enough
to allow $a,b,c$ to cover all their definition domain, i. e. $z < N_{pixel} \times f$ where $N_{pixel}$ is the total number of detector pixels. 

When $m$ in a rational number which means that $\theta$
does not vary with continuity, the summation of terms corresponding to a
proper $\theta$ has to be used in place of the integral.  In particular, for   
$m=1$ (that is for $z\rightarrow \infty$) one value of $\theta$ has
to be considered.

\newpage
\section*{List of figures}

\noindent Fig. 1 Top panel: 3D plot of  bi-dimensional maximum function shape  of a simulated point-source at a distance of 250 cm from the detector plane; bottom panel: on the left the bi-dimensional maximum projected along the shrinking factor axis; on the right its projection along the shrinking base point axis.

\noindent Fig. 2  Left: the beam divergence at finite distance is shown;
right: a simple scheme that summarizes the shrinking procedure; short dashed
line indicates one of possible directions with respect to which source photons will be collimated.

\noindent Fig. 3 A geometrical view of the interpolation in the detector domain due to the shrinking procedure is shown.
The counts recorded by the j-th pixel ($\tilde D(j)$) strongly depend on the neighbouring detector counts ($D(j-1)$, $D(j+1)$ if $m <2$); $a, b, c$ represent the fraction of source flux ``lost" passing through the mask pattern numerical tern  $M( j-1), M(j), M(j+1)$, respectively.

\noindent Fig. 4 Solid line: reconstructed sky image of point source at infinite
    distance (obtained using the standard deconvolution method); dot: focused image of same source simulated at finite distance (say z=250 cm); dashed line: the same for a 5 mm extended source.

\noindent Fig. 5 The left panel shows the image of a point source simulated at
   $z+f=250$ cm  obtained with the standard deconvolution method; the right panel
   shows the image obtained for the same source after the application of the
   shrinking procedure.

\noindent Fig. 6 Reconstructed images for Cd$^{109}$ at 2 meters from the detection
   plane from -40 up to 50 degrees off-axis. The angular size of each bin is 3
 arc minutes (one SuperAGILE sky pixel).

\noindent Fig. 7 Cd$^{109}$ at 2 meters from the detection
   plane at 0 and 30 degrees off-axis in the total field of view of SuperAGILE.

\noindent Fig. 8 SuperAGILE pixel-angle relation (for each of the four detector
    units).

\newpage
\begin{figure}[h!]
\begin{center}
\subfigure{\rotatebox{90}{\includegraphics[width=0.45\textwidth]{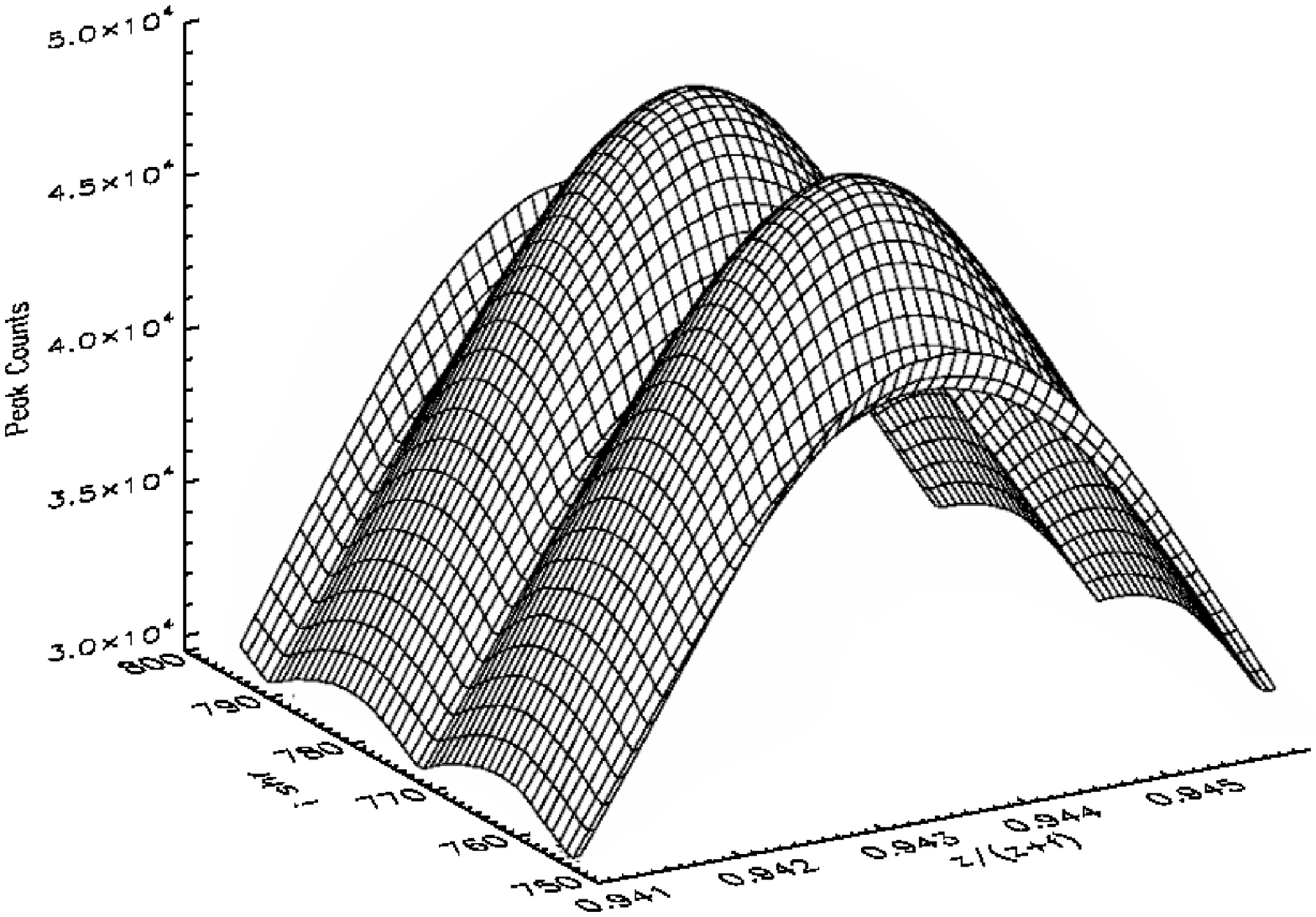}}}
\hspace{0.15cm}
\subfigure{\includegraphics[width=0.4\textwidth]{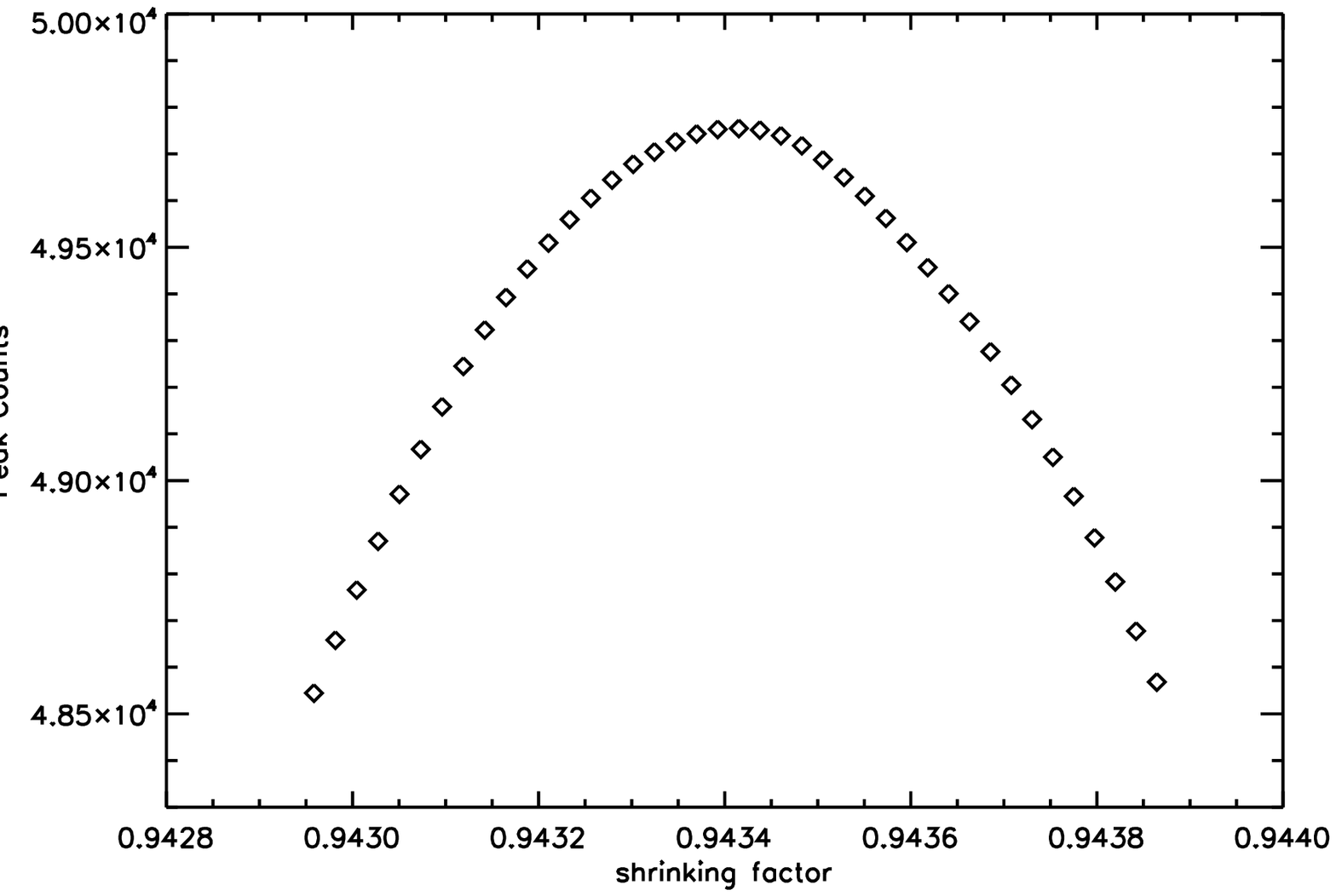}}
\hspace{0.05cm}
\subfigure{\includegraphics[width=0.4\textwidth]{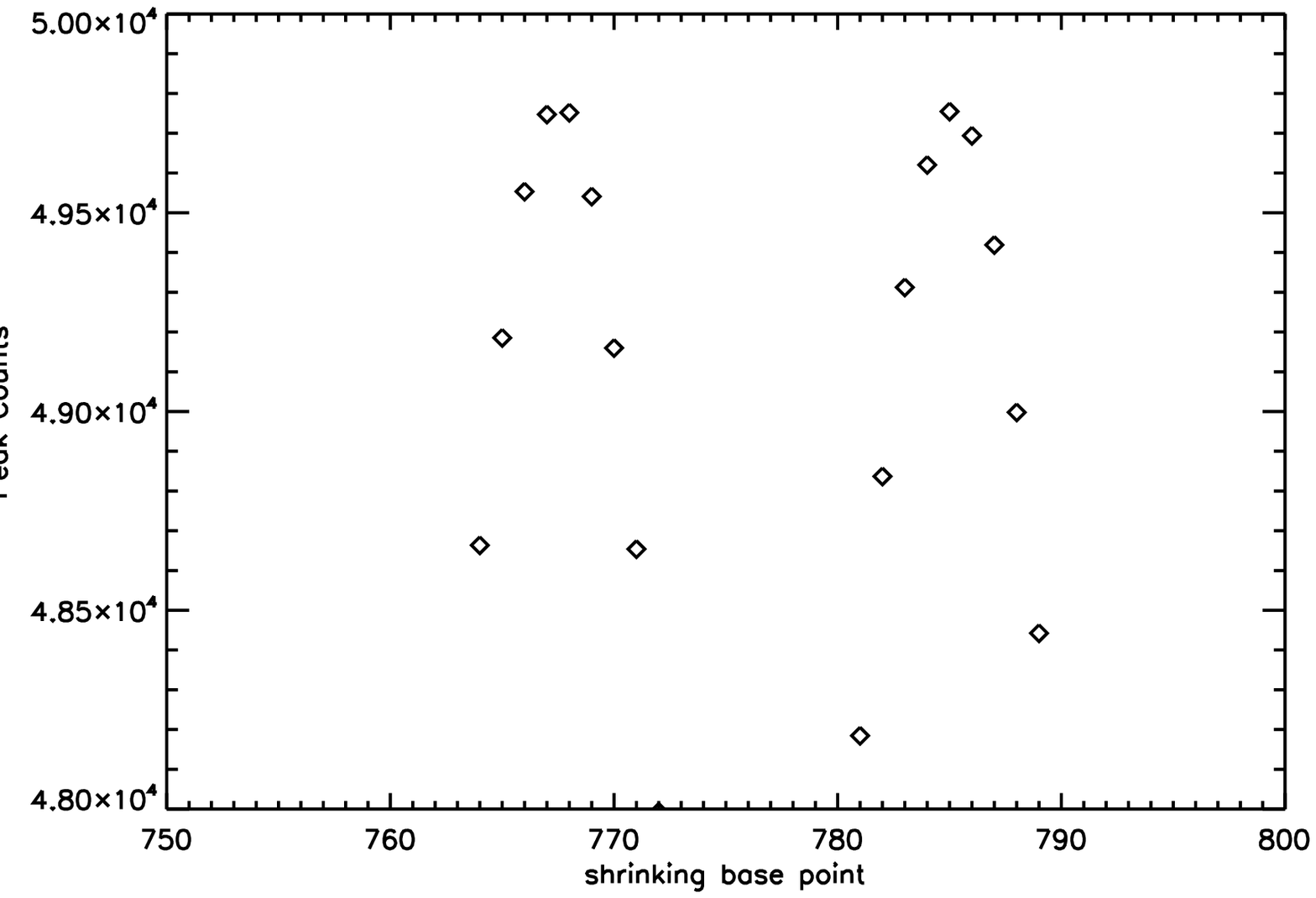}}
\end{center}
  \caption{Top panel: 3D plot of  bi-dimensional maximum function shape  of a simulated point-source at a distance of 250 cm from the detector plane; bottom panel: on the left the bi-dimensional maximum projected along the shrinking factor axis; on the right its projection along the shrinking base point axis.}
 \end{figure}

\newpage

\begin{figure}[h!]
{\includegraphics[width=0.85\textwidth, height=0.65\textwidth]{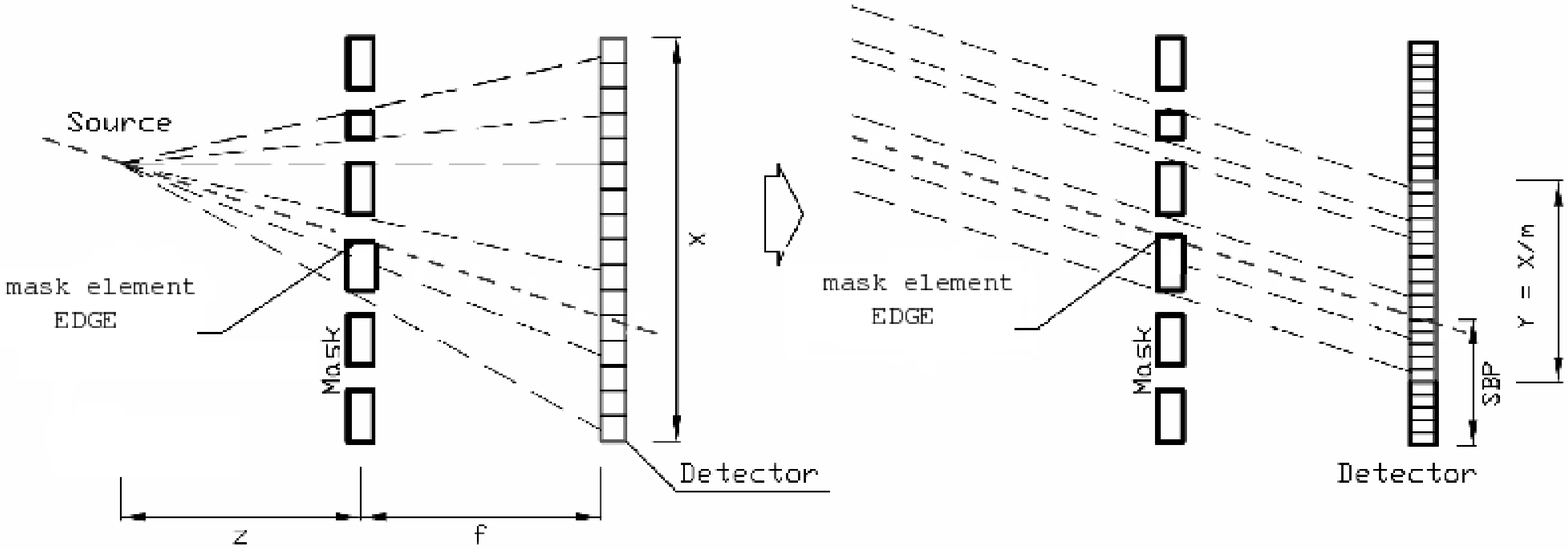}}
\caption{Left: the beam divergence at finite distance is shown; right: a
  simple scheme that summarizes the shrinking procedure; short dashed line
  indicates one of possible directions with respect to which source photons will be collimated.}

\end{figure}

\newpage
\begin{figure}[h!]
\centerline{\includegraphics[width=8.5cm, height=7cm]{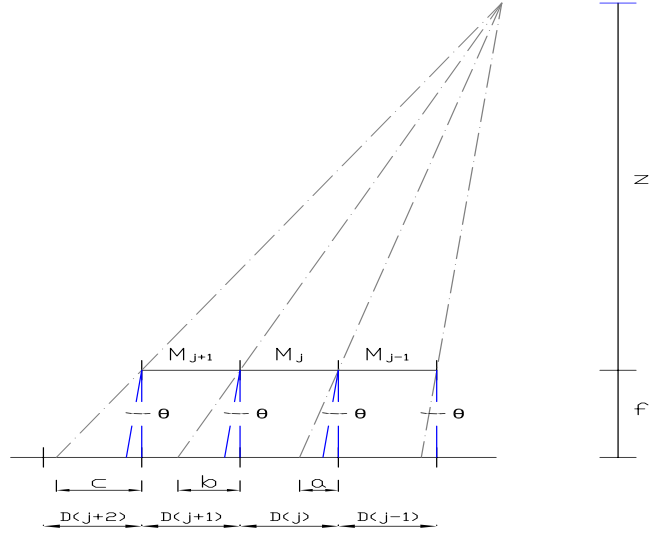}}
  \caption{A geometrical view of the interpolation in the detector domain due to the shrinking procedure is shown.
The counts recorded by the j-th pixel ($\tilde D(j)$) strongly depend on the neighbouring detector counts ($D(j-1)$, $D(j+1)$ if $m <2$); $a, b, c$ represent the fraction of source flux ``lost" passing through the mask pattern numerical tern  $M( j-1), M(j), M(j+1)$, respectively.}
 \end{figure}

\begin{table}[htbp]
\begin{center}
\begin{tabular}{lccc}
\hline
$z+f$ (cm) & peak counts & $\bigtriangleup$f ($\mu$m) \\ 

\hline
\\
100 & 0.84 & 165 \\
130 & 0.83 & 110 \\
150 & 0.83 & 90  \\
180 & 0.83 & 80  \\
200 & 0.82 & 78   \\
230 & 0.82 & 75 \\
250 & 0.82 & 75\\
270 & 0.82 & 60\\
300 & 0.82 & 60\\
\hline
\end{tabular}
\caption{The reconstructed peak counts as a function of source distance to the
  detector plane; third column:  the mask-detector separation error inferred by the statistical error $\bigtriangleup$z as a function of $z+f$.}
\end{center}
\end{table}
\newpage

\begin{figure}[!ht]
\centerline{\includegraphics[width=8cm, height=7.5cm]{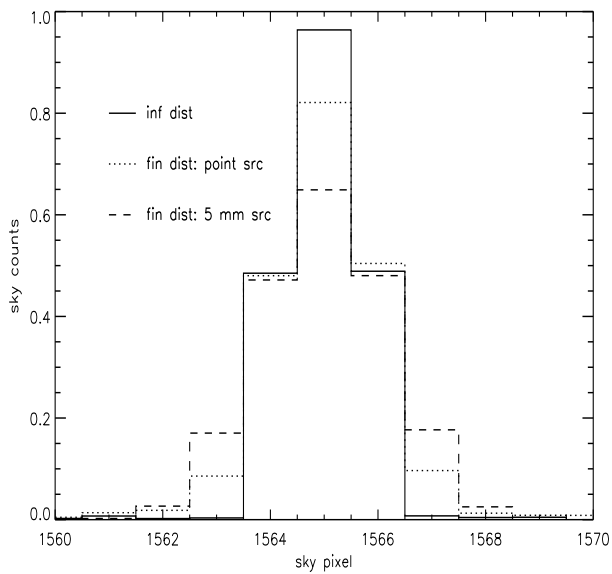}}
  \caption{Solid line: reconstructed sky image of point source at infinite
    distance (obtained using the standard deconvolution method); dot: focused image of same source simulated at finite distance (say z=250 cm); dashed line: the same for a 5 mm extended source.}
 \end{figure}
\newpage

\begin{figure}[ht]
\begin{center}
\subfigure{\includegraphics[width=0.45\textwidth, height= 0.45\textwidth]{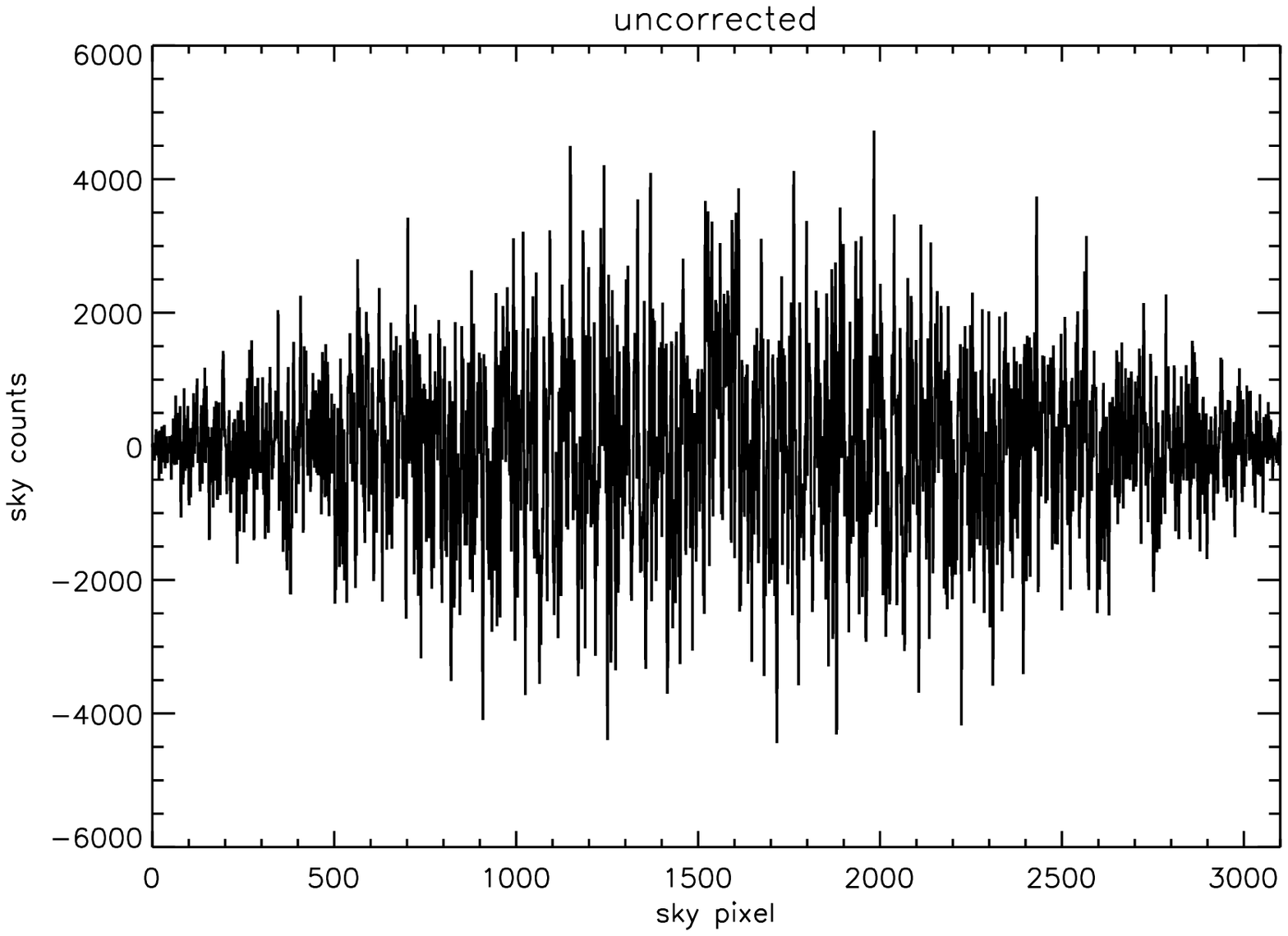}}%
\hspace{0.2cm}%
\subfigure{\includegraphics [width=0.45\textwidth, height= 0.45\textwidth]{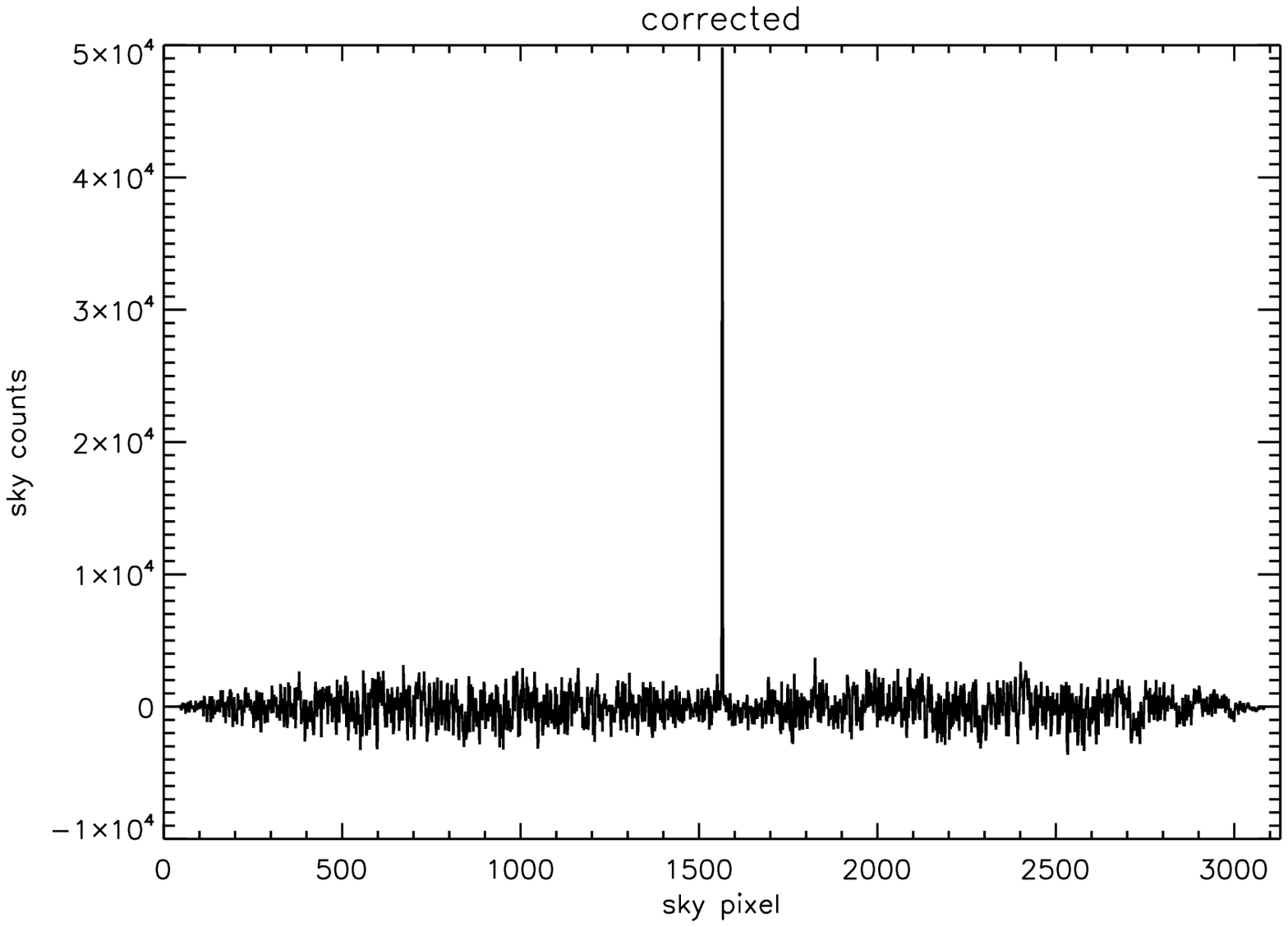}}%
 \caption{The left panel shows the image of a point source simulated at
   $z+f=250$ cm  obtained with the standard deconvolution method; the right panel
   shows the image obtained for the same source after the application of the
   shrinking procedure.}
\end{center}
\end{figure}
\newpage

\begin{figure}[!h]
\begin{center}

\subfigure{\rotatebox{90}{\includegraphics [width=0.25\textwidth]{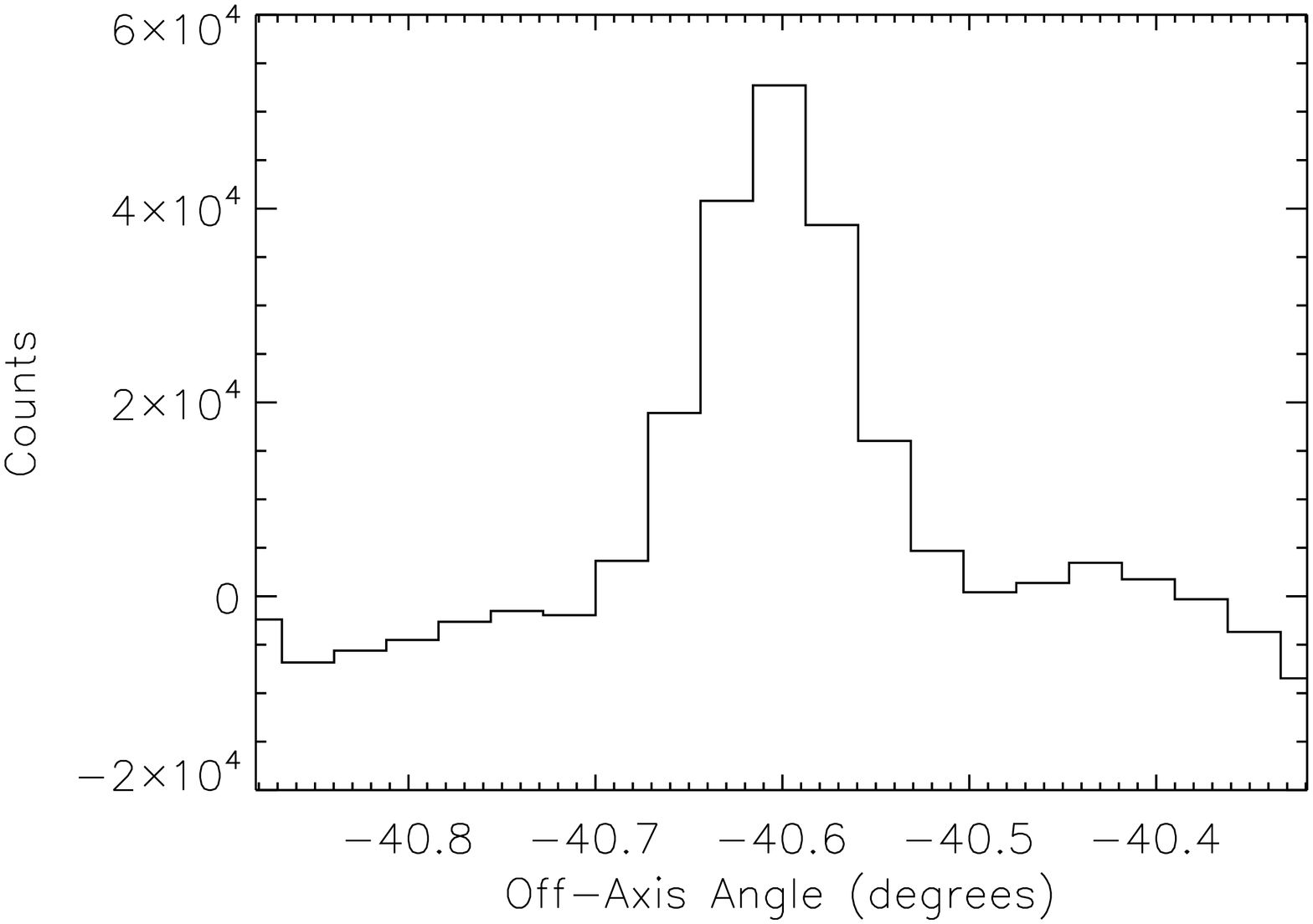}}}%
\hspace{0.35cm}%
\subfigure{\rotatebox{90}{\includegraphics [width=0.25\textwidth]{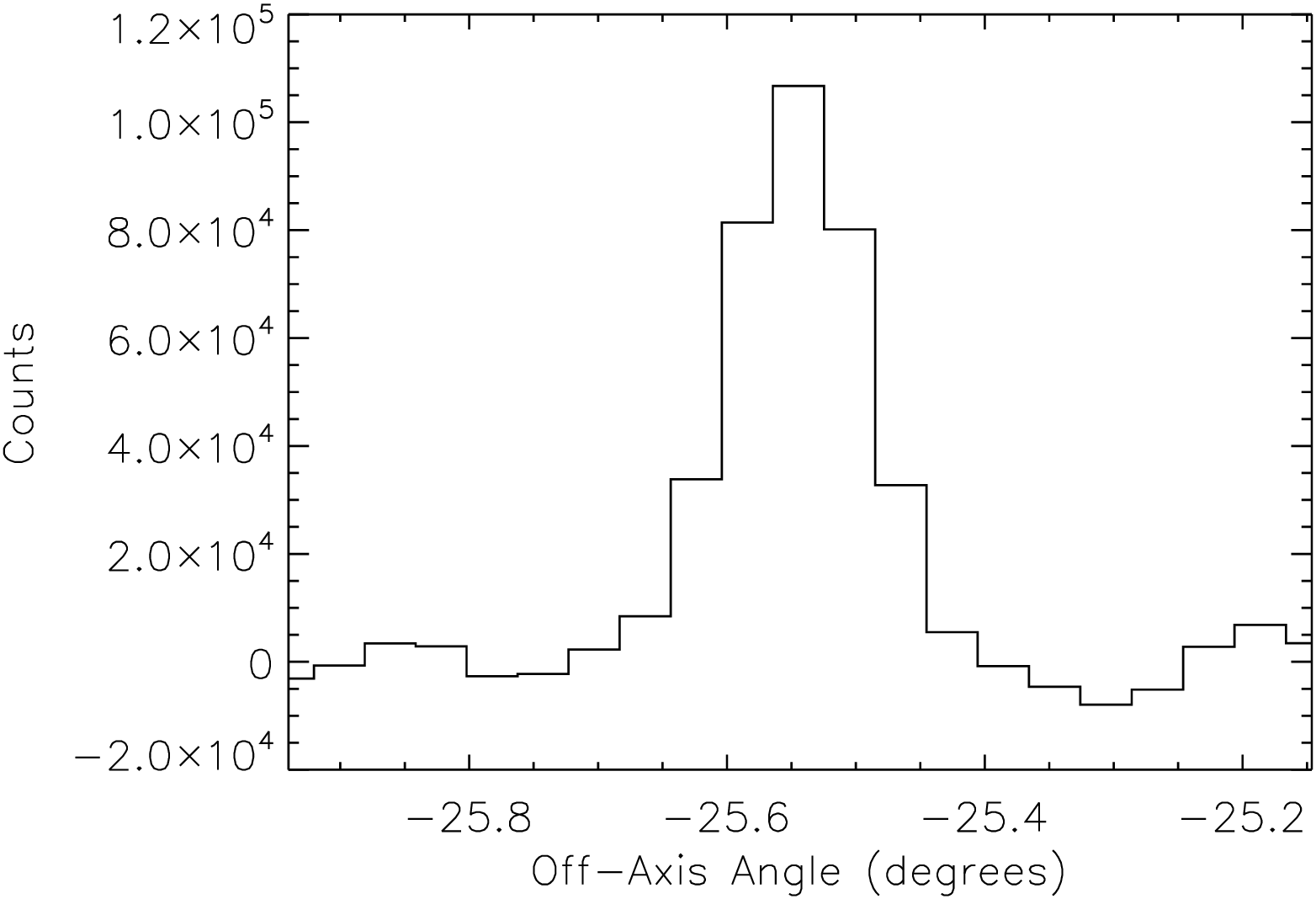}}}%
\hspace{0.25cm}%
\subfigure{\rotatebox{90}{\includegraphics[width=0.25\textwidth]{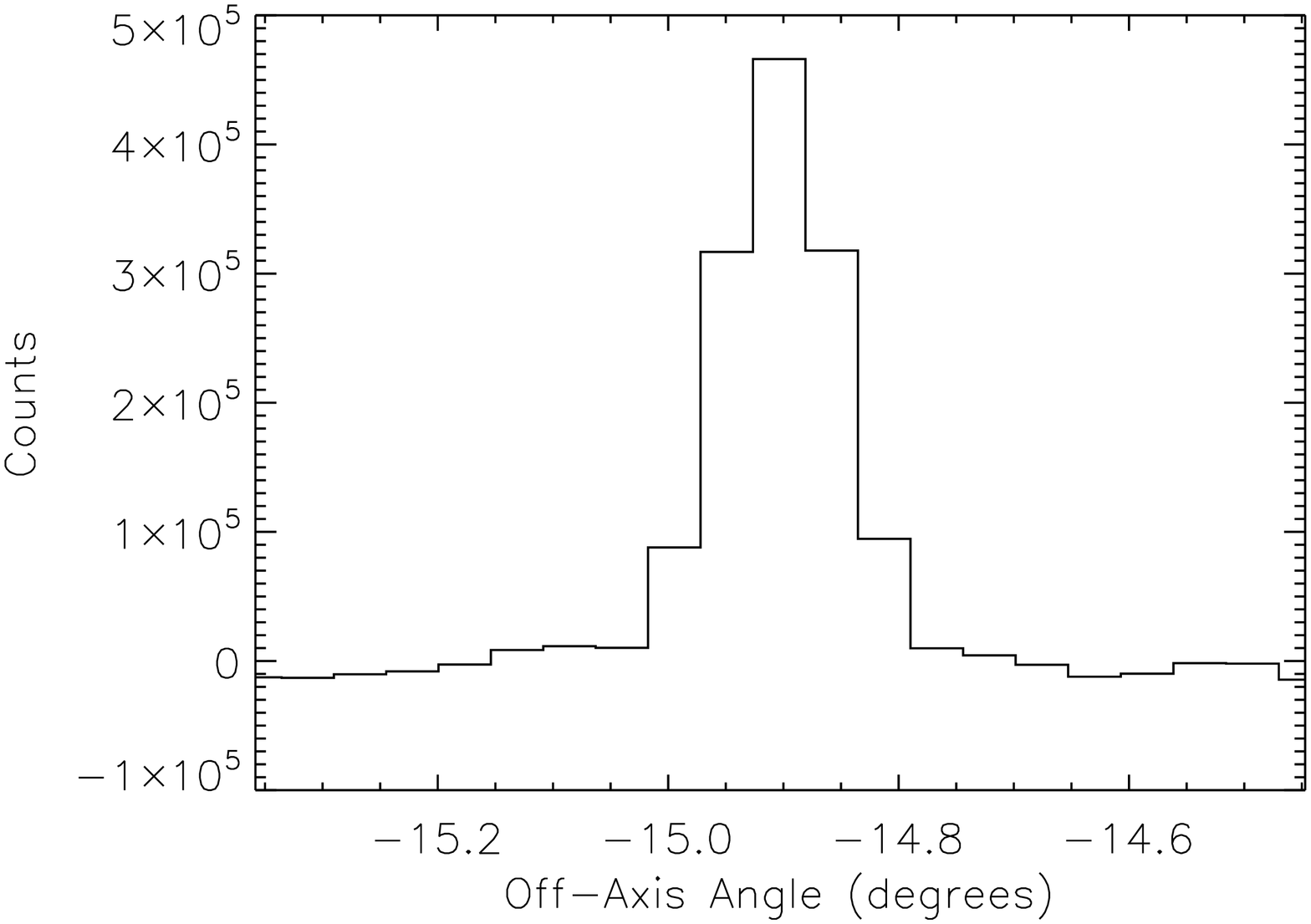}}}%
\hspace{0.2cm}%
\subfigure{\rotatebox{90}{\includegraphics [width=0.25\textwidth]{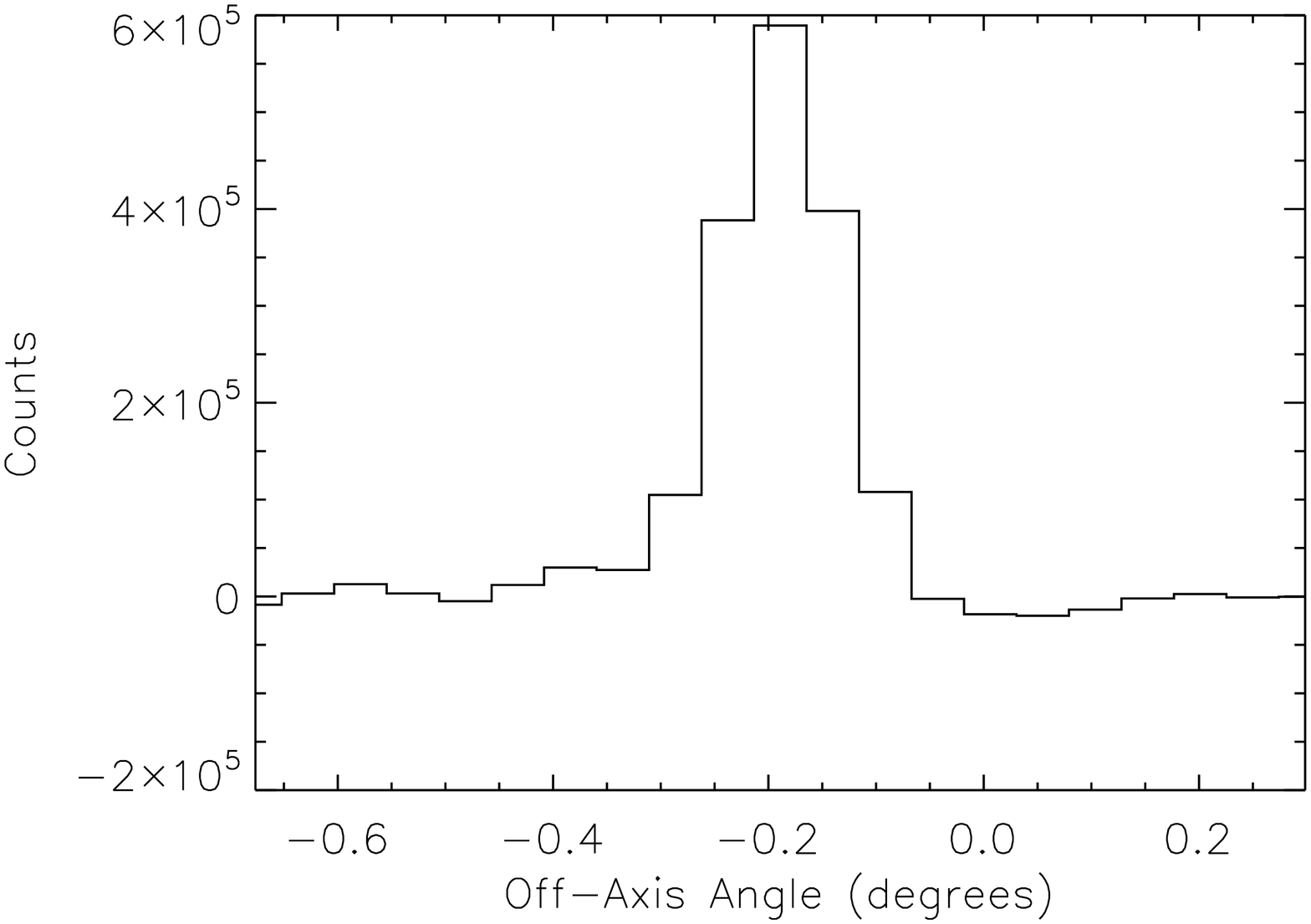}}}%
\hspace{0.3cm}%
\subfigure{\rotatebox{90}{\includegraphics[width=0.25\textwidth]{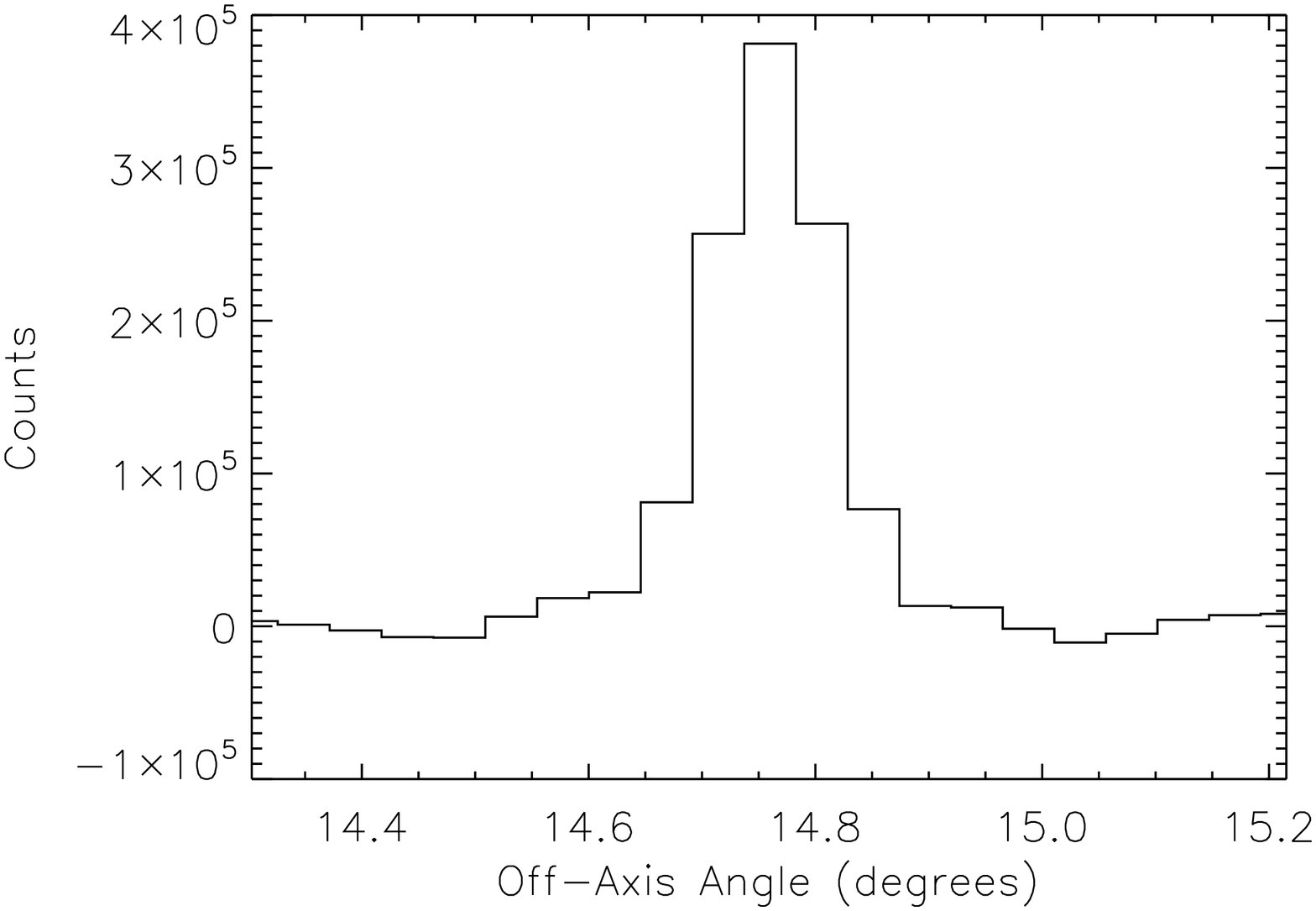}}}%
\hspace{0.35cm}%
\subfigure{\rotatebox{90}{\includegraphics [width=0.25\textwidth]{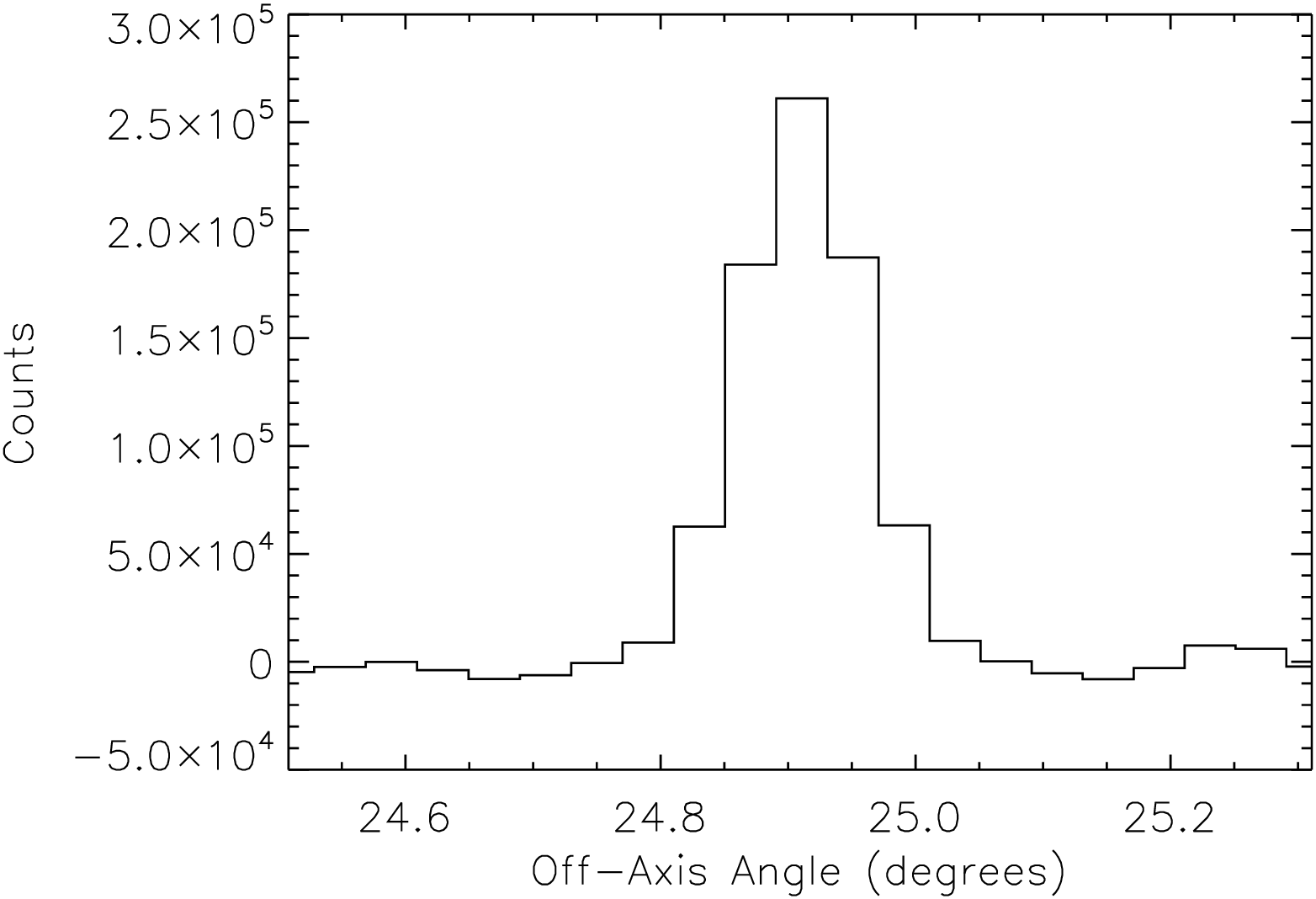}}}%
\hspace{0.35cm}%
\subfigure{\rotatebox{90}{\includegraphics [width=0.25\textwidth]{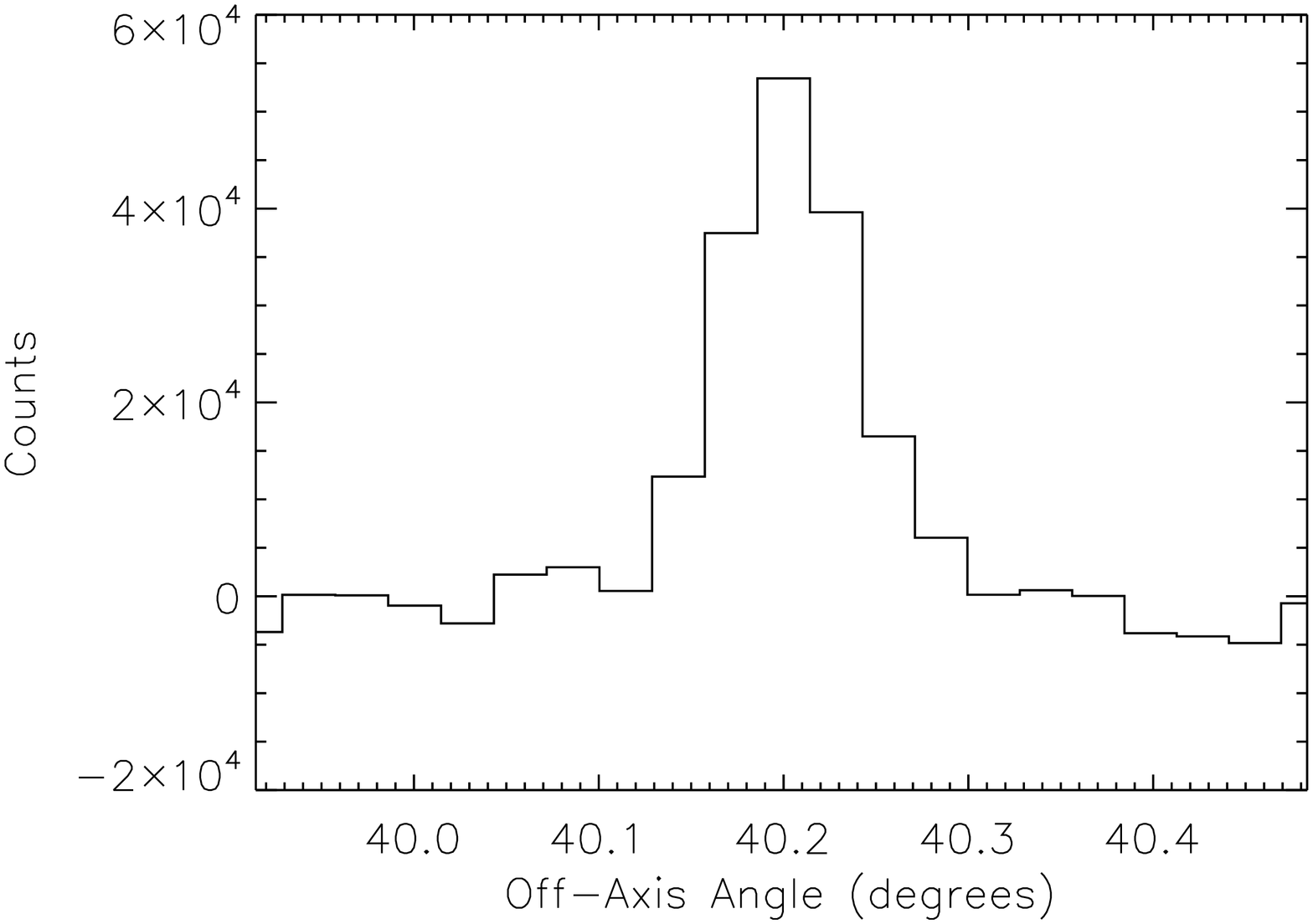}}}%
\hspace{0.35cm}%
\subfigure{\rotatebox{90}{\includegraphics [width=0.25\textwidth]{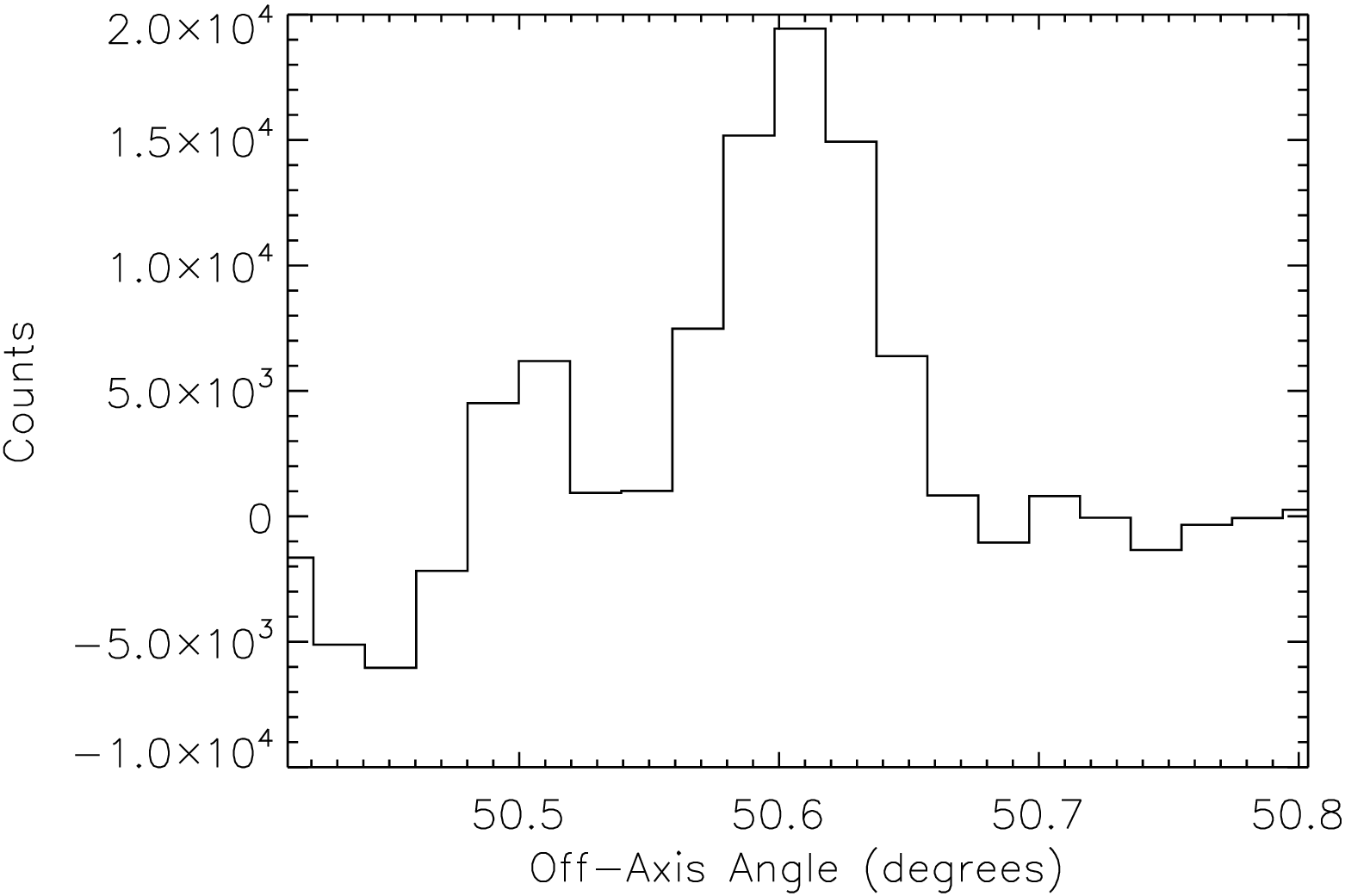}}}%
\end{center}
 \caption{Reconstructed images for Cd$^{109}$ at 2 meters from the detection
   plane from -40 up to 50 degrees off-axis. The angular size of each bin is 3
 arc minutes (one SuperAGILE sky pixel).}

\end{figure}
\newpage

\begin{figure}[!ht]
\begin{center}
\subfigure{\rotatebox{90}{\includegraphics [width=0.45\textwidth, height=0.45\textwidth]{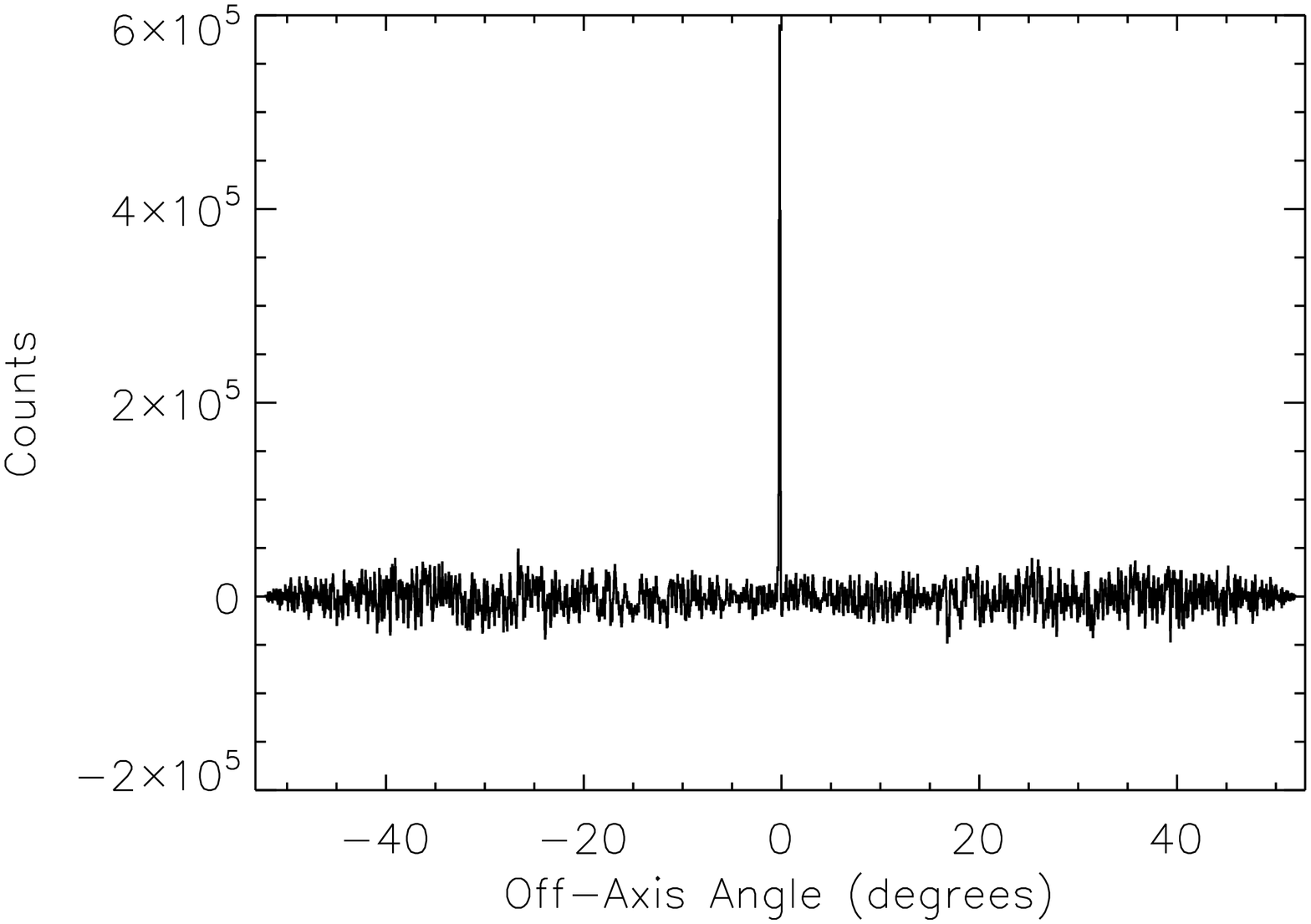}}}%
\hspace{0.45cm}%
\subfigure{\rotatebox{90}{\includegraphics [width=0.45\textwidth, height=0.45\textwidth]{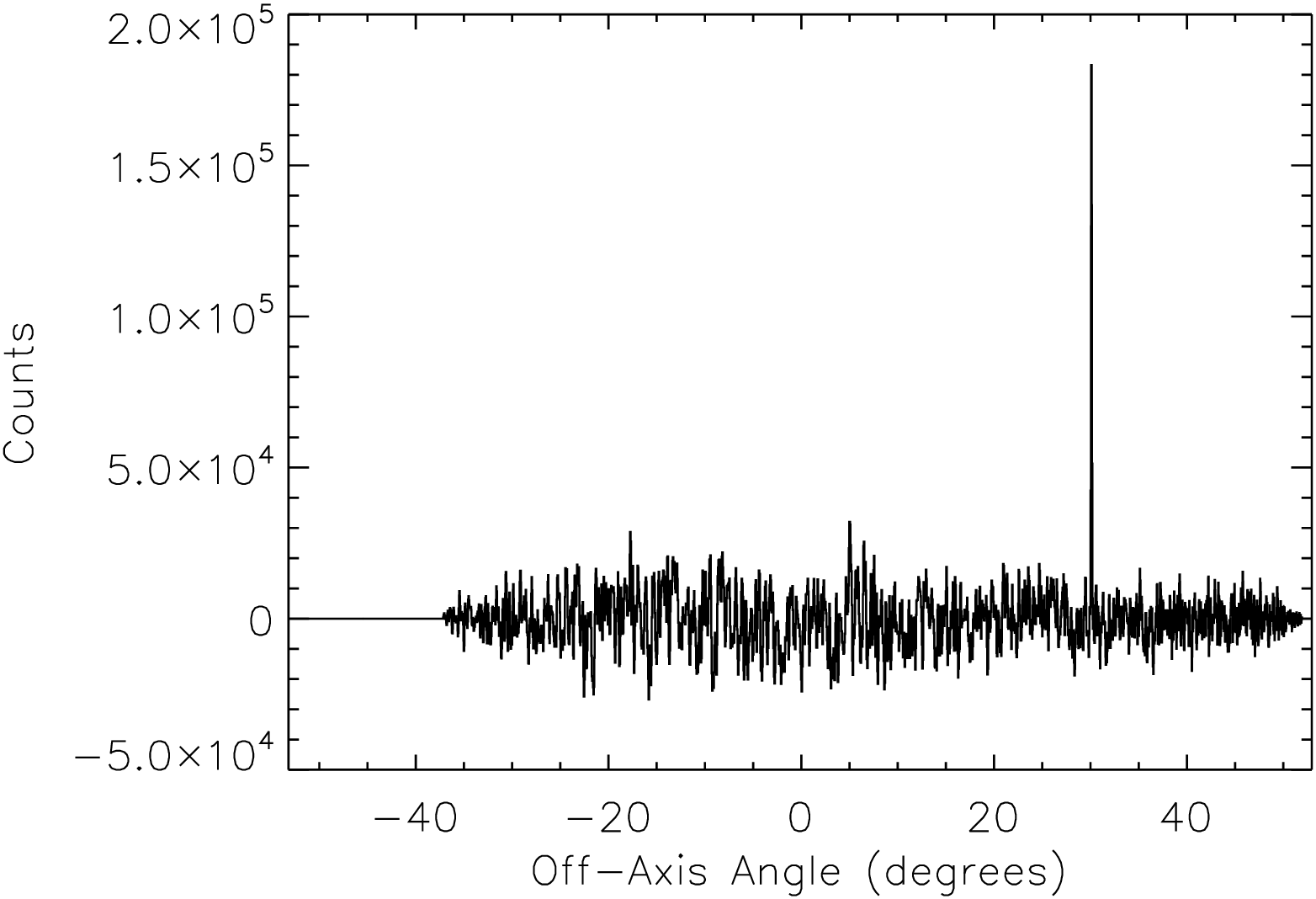}}}%

\end{center}

 \caption{Cd$^{109}$ at 2 meters from the detection
   plane at 0 and 30 degrees off-axis in the total field of view of SuperAGILE. }
\end{figure}
\newpage

\begin{figure}[!ht]

\begin{center}
\subfigure{\rotatebox{90}{\includegraphics [width=0.35\textwidth]{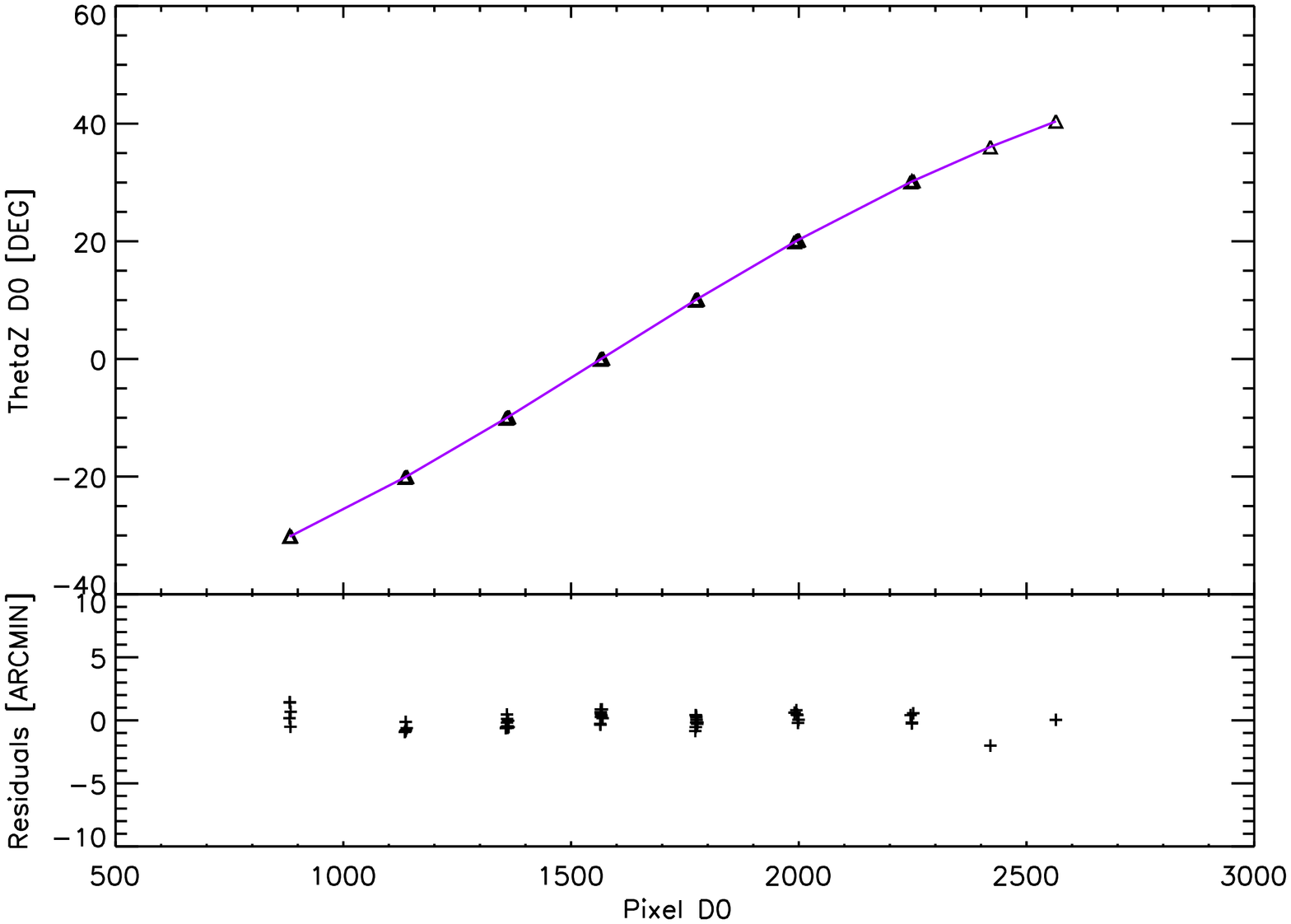}}}
\hspace{0.0cm}%
\subfigure{\rotatebox{90}{\includegraphics [width=0.35\textwidth]{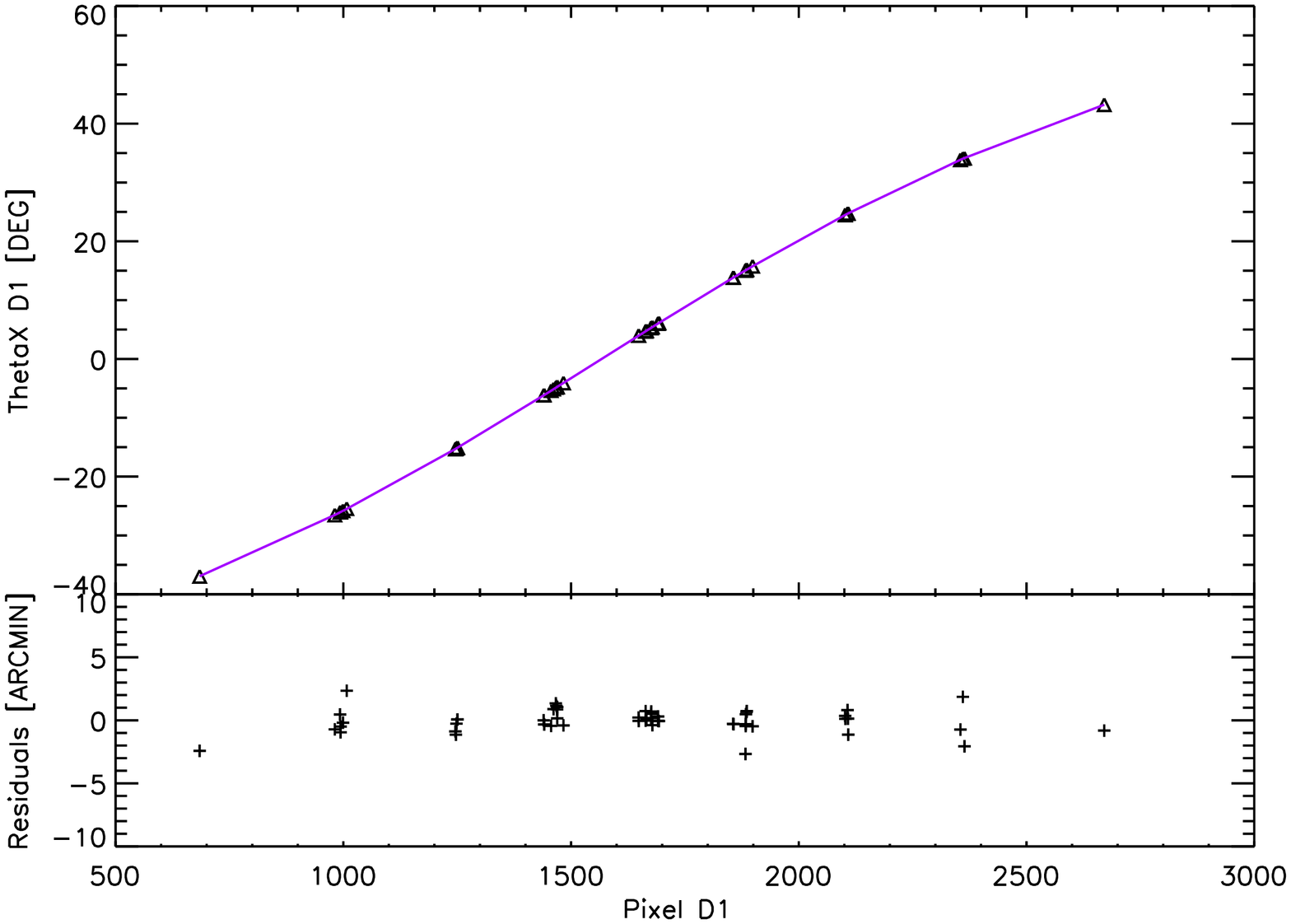}}}
\hspace{0.15cm}%
\subfigure{\rotatebox{90}{\includegraphics [width=0.35\textwidth]{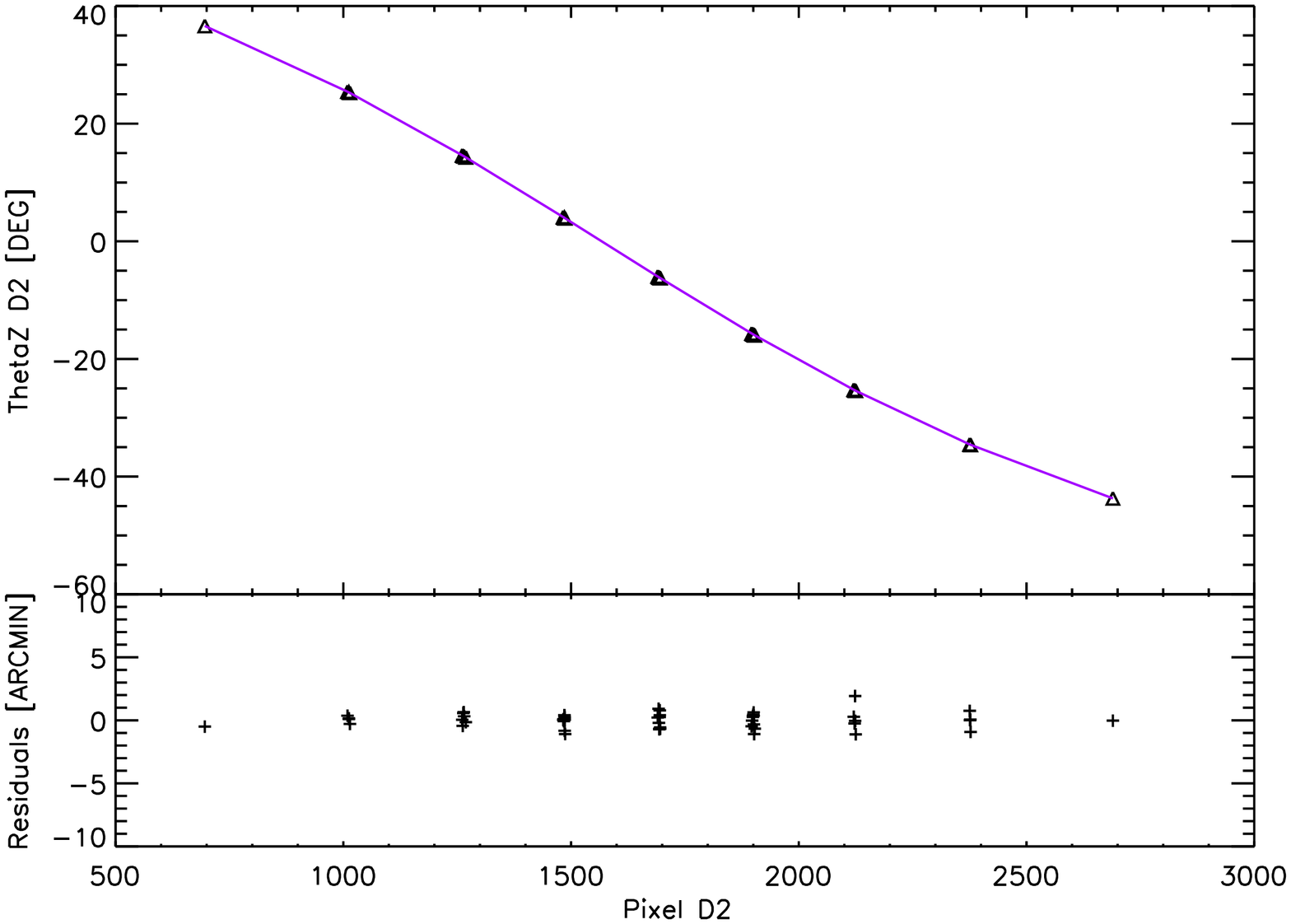}}}
\hspace{0.15cm}%
\subfigure{\rotatebox{90}{\includegraphics [width=0.35\textwidth]{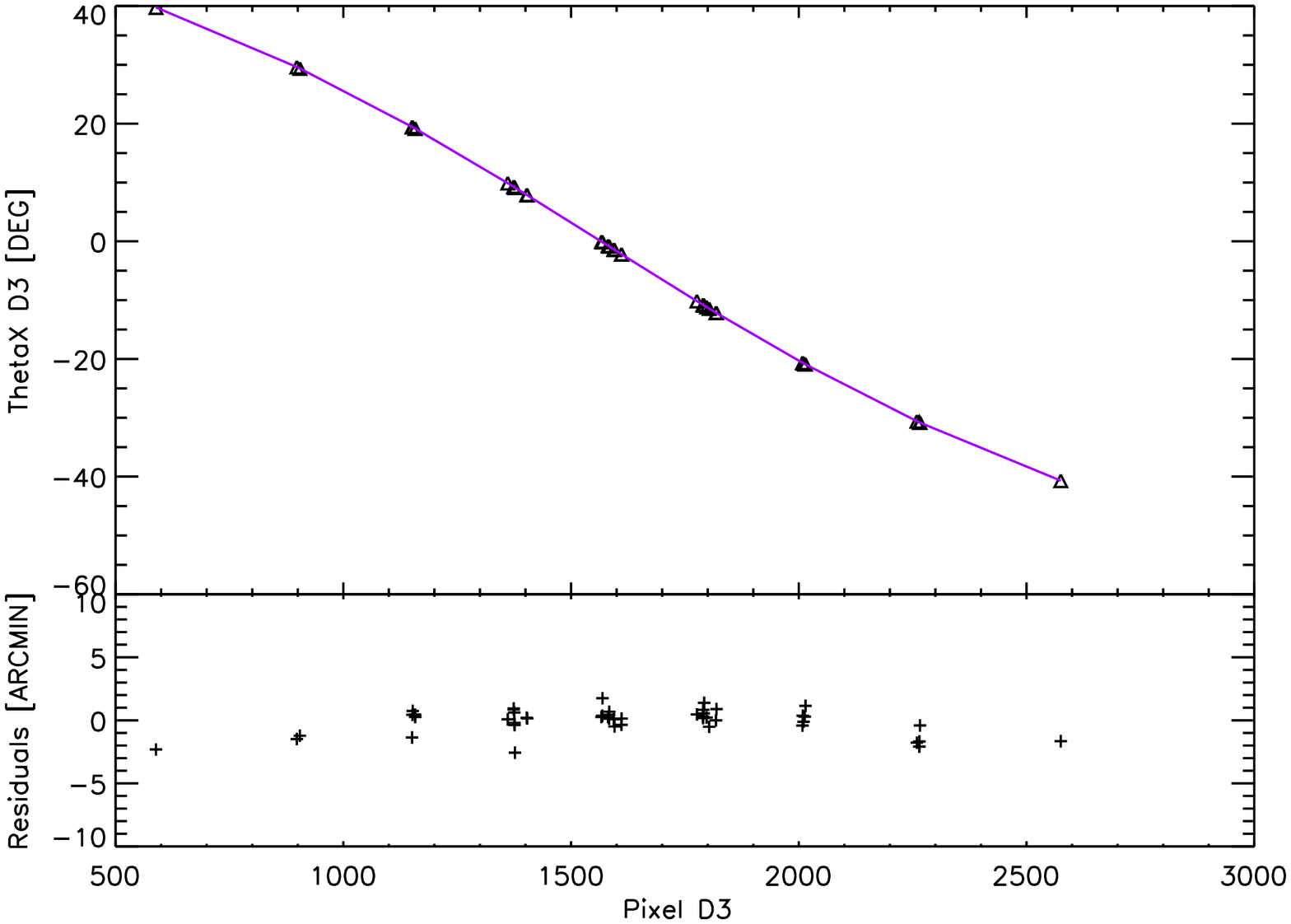}}}
\end{center}
  \caption{SuperAGILE pixel-angle relation (for each of the four detector
    units). }

 \end{figure}

\end{document}